\documentclass[a4paper,fleqn]{cas-sc}
\usepackage{cuted}
\usepackage{pifont}
\usepackage{tabularx}  
\usepackage{placeins}

\usepackage[super,sort&compress]{natbib}
\setcitestyle{comma,numbers,super,open={},close={}} 
\makeatletter
\renewcommand\@biblabel[1]{#1.}
\makeatother
\usepackage{upgreek}
\usepackage{indentfirst} 
\usepackage{amsmath}
\usepackage{bm}
\usepackage{lineno}
\usepackage{xcolor}
\usepackage{underscore}
\usepackage{lineno}

\def\tsc#1{\csdef{#1}{\textsc{\lowercase{#1}}\xspace}}
\tsc{WGM}
\tsc{QE}
\tsc{EP}
\tsc{PMS}
\tsc{BEC}
\tsc{DE}
\begin{document}
\let\WriteBookmarks\relax
\def\floatpagepagefraction{1}
\def\textpagefraction{.001}
\shorttitle{}

% Short author
\shortauthors{Pan et~al.}

% Main title of the paper
\title [mode = title]{An analytical framework to assess static versus dynamic triggering of fault-slip rockbursts}

% \tnotetext[1]{This document is the results of the research
%    project funded by the National Science Foundation.}

% \tnotetext[2]{The second title footnote which is a longer text matter
%    to fill through the whole text width and overflow into
%    another line in the footnotes area of the first page.}

\author[1,2,3]{Wenbo Pan}
\credit{Methodology, Software, Validation, Visualization, Data curation, Formal Analysis, Investigation, Writing - Original draft}

\author[1]{Zixin Zhang}
\credit{Resources, Supervision, Writing - Review \& Editing, Project administration, Funding Acquisition}

\author[2,3]{Qinghua Lei}[orcid = 0000-0002-3990-4707]
\cormark[1]
\ead{qinghua.lei@geo.uu.se}
\credit{Conceptualization, Methodology, Investigation, Resources, Supervision, Writing - Review \& Editing, Funding Acquisition}

\affiliation[1]{organization={Department of Geotechnical Engineering, College of Civil Engineering, Tongji University},
                city={Shanghai}, 
                country={China}}
\affiliation[2]{organization={State Key Laboratory of Intelligent Coal Mining and Strata Control},
                city={Beijing},
                country={China}}

\affiliation[3]{organization={Department of Earth Sciences, Uppsala University},
                city={Uppsala},
                country={Sweden}}

% Corresponding author text
\cortext[cor1]{Corresponding author}
% \cortext[cor2]{Principal corresponding author}
% \fntext[fn1]{This is the first author footnote, but is common to third
  % author as well.}
% \fntext[fn2]{Another author footnote, this is a very long footnote and
%   it should be a really long footnote. But this footnote is not yet
%   sufficiently long enough to make two lines of footnote text.}

% \nonumnote{This note has no numbers. In this work we demonstrate $a_b$
%   the formation Y\_1 of a new type of polariton on the interface
%   between a cuprous oxide slab and a polystyrene micro-sphere placed
%   on the slab.
%   }

% Here goes the abstract

\begin{abstract}
Fault-slip rockbursts, triggered by seismic rupture of nearby or remote faults, constitute a significant geohazard during deep underground excavations. Although these events occur frequently in underground projects, their underlying mechanisms are not yet fully understood. Most studies tacitly assume dynamic stress waves as the main triggering factor, often disregarding  the role of coseismic static stress changes associated with fault slip. This paper introduces a novel analytical framework to diagnose both static and dynamic coseismic stress perturbations and quantify their contributions to fault-slip rockburst around a circular tunnel. Building on linear elastic fracture mechanics, seismic source theory, and the Kirsch solution, the model assesses whether coseismically elevated maximum tangential stress on the tunnel boundary under static and dynamic triggering effects is sufficient to induce failure around the tunnel. We extensively test our framework using synthetic case studies that represent typical fault-slip rockburst scenarios. Our results yield a rockburst hazard map that delineates regions of elevated triggering potential in the near-field and far-field of the seismogenic fault, and classify the triggering types as static, dynamic, or dual. We perform a comprehensive parametric sensitivity analysis to investigate how key factors, including seismic source characteristics, rock mass properties, and in-situ stress conditions, influence the spatial distribution of rockburst susceptibility. The model is further applied to a historical fault-slip rockburst event at the Gotthard Base Tunnel, effectively capturing the triggering mechanism of the observed failure. Our research provides a physically grounded and computationally efficient analytical framework with the results carrying significant implications  for rockburst hazard assessments during deep underground excavations.

% \noindent\texttt{\textbackslash begin{abstract}} \dots 
% \texttt{\textbackslash end{abstract}} and
% \verb+\begin{keyword}+ \verb+...+ \verb+\end{keyword}+ 
% which
% contain the abstract and keywords respectively. 

% \noindent Each keyword shall be separated by a \verb+\sep+ command.
\end{abstract}

% \begin{graphicalabstract}
% \includegraphics{figs/cas-grabs.pdf}
% \end{graphicalabstract}


\maketitle

\section{Introduction}

Deep underground excavations in hard rock under high in-situ stress conditions are frequently encountered with rockburst events \cite{ortlepp1994rockburst,kaiser2012design,li2012situ,li2019discussions,li2021principles,askaripour2022rockburst,he2023review}, which pose serious risks to the safety of mining and construction operations. These rockburst events, increasingly documented in deep mining and tunneling projects worldwide \cite{ma2015rockburst,keneti2018review,rehbock2018fault,li2021principles,askaripour2022rockburst,gong2023strength,he2023review}, are characterized by a sudden release of stored elastic strain energy in the rock mass, leading to violent ejection of rock fragments \cite{kaiser2012design,li2021principles,he2023review}. A particularly hazardous type, fault-slip rockburst, is triggered by stress perturbations induced by the rupture of a nearby or remote seismogenic fault, with the hypocenter typically located tens to hundreds of meters away from the excavation site \cite{simser2001geotechnical,kaiser2012design,stacey2016addressing,rehbock2018fault,li2021principles}. These perturbations can generate stress concentrations intense enough to initiate rock failure around the tunnel boundary \cite{manouchehrian2017analysis,li2017failure,su2017experimental,hu2018experimental,gong2023strength}. Fault-slip rockbursts, where the source of the seismic event and the resulting damage occur at different locations, differ fundamentally from strainbursts, where the source and damage are at the same location \cite{stacey2016addressing,li2019discussions,he2023review}.

Over the past years, substantial research efforts have been dedicated to investigating the mechanisms of fault-slip rockbursts based on experimental and numerical approaches \cite{kaiser2013critical,wang2017numerical,su2017experimental,hu2018experimental,cai2021fault,gao2021numerical,cao2023numerical,ren2023characteristics}. Fault-slip rockbursts were traditionally understood to result from the combined effects of radiated seismic waves, excavation-induced stress concentrations, and geological discontinuities \cite{wang2017numerical,manouchehrian2017analysis,hu2018experimental,gao2021numerical,he2023review}. In particular, fault-slip rockbursts were typically considered to be caused by the superposition of excavation-induced high static stresses and fault rupture-generated dynamic stresses \cite{hu2018experimental,cai2021fault,cao2023numerical,ren2023characteristics}. Several studies have demonstrated that once the surrounding rock reaches a critically stressed state due to excavation, even low-amplitude seismic waves from remote fault slip events may be sufficient to trigger severe rockbursts\cite{su2017experimental,hu2018experimental,su2024influence}. While the dominant role of coseismic dynamic triggering associated with seismic waves has been widely recognized, the potential contribution of coseismic static stress changes remains almost underexplored. These static stress changes arise from permanent stress redistribution induced by fault slip and are most pronounced in the near-field region, typically defined as within one to two rupture lengths of the seismogenic fault \cite{king1994static,freed2005earthquake,velasco2008global}. Such static stress changes decay rapidly with distance from the fault, becoming negligible in the far-field region beyond the near-field zone.

Extensive seismological studies have shown that coseismic static stress changes play a crucial role in aftershock triggering \cite{king1994static,hardebeck1998static}, especially in near-field regions, highlighting the competing roles of static versus dynamic triggering effects in aftershock activity \cite{cotton1997dynamic,freed2005earthquake}. Fault-slip rockbursts share fundamental physical characteristics with aftershocks. On the one hand, both are fault-related phenomena originating from rupture along a seismogenic fault and initiating at locally reactivated zones. On the other hand, both are governed by similar controlling factors, including the pre-earthquake stress state, earthquake-induced stress changes, and the mechanical strength of rock masses \cite{he2023review,pan2023earthquake,pan2024slip,palgunadi2024rupture,gabriel2024fault,pan49703313d}. It is thus natural to anticipate that fault-slip rockbursts may be coseismically triggered by both static and dynamic effects. It is important to emphasize that, in this study, both static and dynamic effects are considered "coseismic", that is, they occur contemporaneously with the seismic event and are directly induced by fault rupture. This differs from conventional interpretations in rockburst theory, where "static" effects are attributed to excavation-induced stress concentrations rather than coseismic processes \cite{he2017rock,li2017failure,manouchehrian2017analysis,cai2020monitoring,cai2021fault,ren2023characteristics}.

The objective of this study is to establish a novel analytical framework that incorporates both static and dynamic triggering effects to investigate the triggering mechanisms of fault-slip rockbursts, with a particular focus on clarifying the interplay and competition of coseismic static stress changes and dynamic stress waves. The  analytical formulation developed further provides a parsimonious and robust framework for risk assessment of fault-slip rockbursts during deep underground excavations. The remainder of this paper is organized as follows. Section \ref{section:analytical_model} presents the detailed formulation and derivation of our analytical model. Section \ref{section:analytical-results} reports the analytical calculation results for both static and dynamic triggering effects under a representative fault-slip rockburst scenario. Section \ref{section:application} applies our model to a documented historical rockburst event at the Gotthard Base Tunnel in Switzerland. Finally, in Section \ref{section:conclusions}, a discussion on rockburst mechanisms is provided and some conclusions are drawn.

\section{Analytical framework}
\label{section:analytical_model}

\begin{figure*}
	\centering
	\includegraphics[width=.9\textwidth]{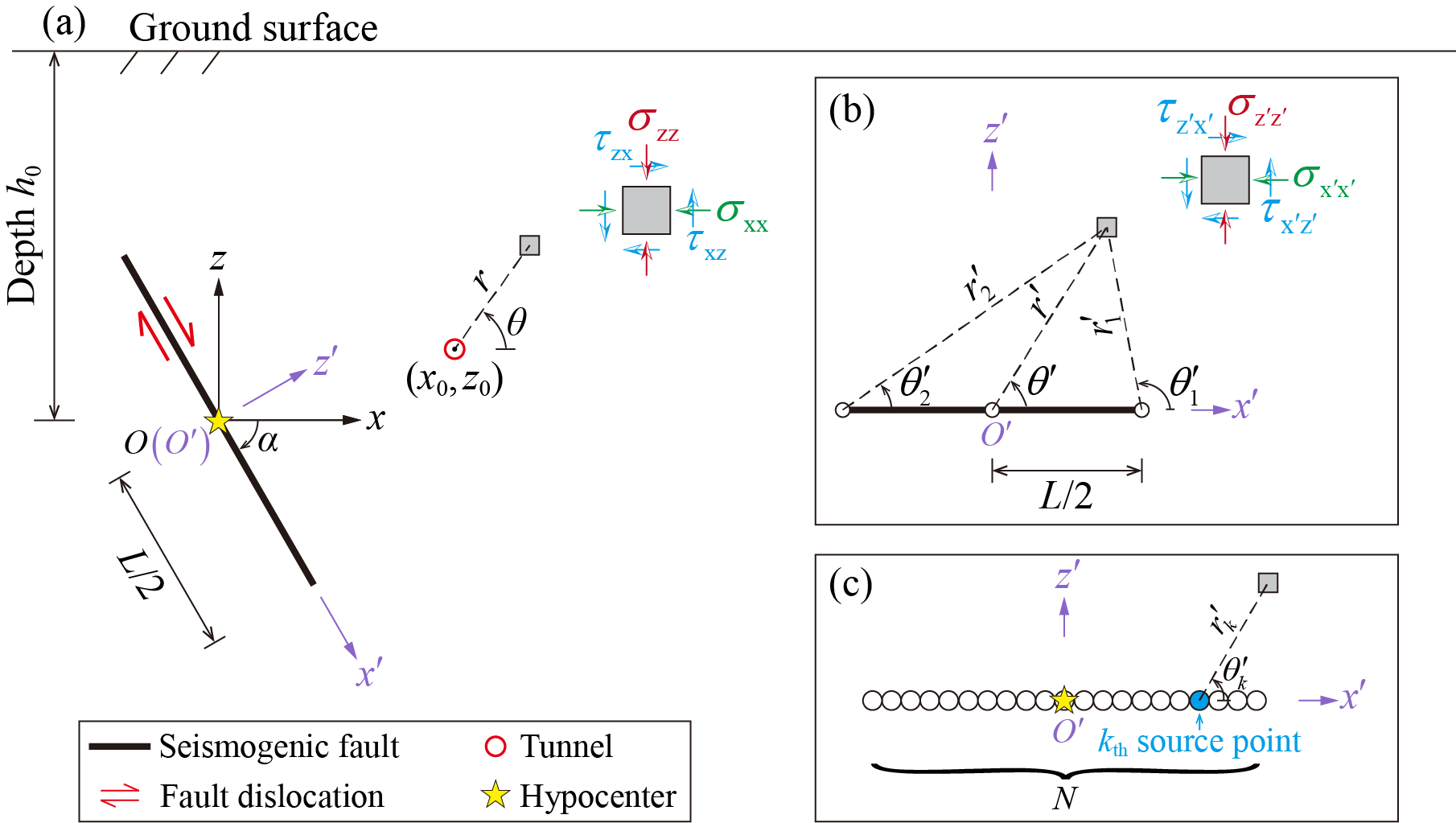}
	\caption{(a) Schematic of the 2D model used to analyze static and dynamic stress changes induced by fault slip in the rock mass where a tunnel is situated. Both the global coordinate system \((x,\, z)\) and the local coordinate system \((x',\, z')\) are defined with their origin at the fault center; the $x'$-axis and $z'$-axis are oriented along and perpendicular to the fault-dip direction, respectively. (b) Definition of the tri-polar coordinate system used for analyzing coseismic static stress redistribution. The primary polar coordinates ($r'$,$\theta'$) originate from the fault center, while two additional polar systems, \((r_1',\, \theta_1')\) and \((r_2',\, \theta_2')\), originate from the right-hand and left-hand fault tips, respectively. (c) Discretizations of the seismogenic fault into $N$ segments, each represented by a source point. The local polar coordinate system for the $k_{\rm{th}}$ source point is $(r_{k}^\prime,\theta_{k}^\prime)$ with the origin at the source point.}
	\label{FIG:SchematicModel}
\end{figure*}

Fault-slip rockbursts are driven by stress concentrations in rock masses surrounding the excavation, which result from disturbances induced by coseismic static and dynamic stress changes associated with seismogenic fault rupture. In our work, the linear elastic fracture mechanics theory is used for evaluating the coseismic static stress changes induced by fault slip \cite{atkinson2015fracture}. Coseismic dynamic stress changes are quantified using the seismic source theory \cite{stein2009introduction}, wherein the seismogenic fault are conceptualized as an assembly of point sources \cite{lei2022reply}. The classical Kirsch solution is also employed for calculating the stress field around a circular tunnel embedded in an infinite elastic medium subjected to far-field stresses \cite{jaeger2009fundamentals}. These classical theories serve as the key building blocks of our analytical framework for analyzing fault-slip rockbursts. Here, our framework is formulated in a two-dimensional (2D) plane strain setting, as shown in Fig.~\ref{FIG:SchematicModel}, representative of tunneling in subsurface rock.

\subsection{Coseismic static stress variation}
\label{section:coseismic_static_solution}

The redistribution of static stress field induced by fault slip is modeled based on the linear elastic fracture mechanics theory \cite{atkinson2015fracture}. In the local coordinate system $(x',\, z')$, with the origin placed at the midpoint of the fault (Fig.~\ref{FIG:SchematicModel}b), the coseismic static stress tensor $\boldsymbol{\sigma}_{i'j'}^{\text{st}}$ is written as:
\begin{align}
    \boldsymbol{\sigma}_{i'j'}^{\text{st}} &= 
    \begin{bmatrix}
        \sigma_{x'x'}^{\text{st}} & \tau_{x'z'}^{\text{st}} \\
        \tau_{z'x'}^{\text{st}} & \sigma_{z'z'}^{\text{st}}
    \end{bmatrix}, \label{eq:coseismic_static_stress_tensor} \\
    \sigma_{x'x'}^{\text{st}} &= \sigma_{x'x'}^{\text{ini}} + \Delta \sigma_\mathrm{II} 
    \left[ 2r' R^{-1} \sin(\theta' - \Theta)  - (L/2)^2 r' R^{-3} \sin \theta' \cos 3\Theta \right], \notag \\
    \sigma _{z'z'}^{{\rm{st}}} &= \sigma _{z'z'}^{{\rm{ini}}} + \Delta {\sigma _\mathrm{II}}\left( {{(L/2)^2}r'{R^{ - 3}}\sin \theta '\cos 3\Theta } \right), \notag \\
    \tau_{x'z'}^{\text{st}} &= \tau_{z'x'}^{\text{st}} = \tau_{x'z'}^{\text{ini}} + \Delta \sigma_\mathrm{II} \left[
    r' R^{-1} \cos(\theta' - \Theta) - 1- (L/2)^2 r' R^{-3} \sin \theta' \sin 3\Theta \right], \notag \\
    R &= {\left( {{{r'}_1}{{r'}_2}} \right)^{1/2}}, \Theta = \left( {{{\theta '}_1} + {{\theta '}_2}} \right)/2, \notag
\end{align}
\noindent where we adopt the geomechanics sign convention with compressive stresses being positive, and the superscript `st' denotes the stress changes due to coseismic static effects. The geometric variables such as $\theta^{\prime}$, $\theta_1^{\prime}$, $\theta_2^{\prime}$, $r^{\prime}$, $r_1^{\prime}$,$ r_2^{\prime}$ are defined in Fig.~\ref{FIG:SchematicModel}b. The initial far-field stress tensor in the local $(x',\,z')$ coordinate system, $\boldsymbol{\sigma}_{i'j'}^{\text{ini}}$, is obtained by rotating the initial far-field stress tensor $\boldsymbol{\sigma}_{ij}^{\text{ini}}$ in the global $(x,\,z)$ coordinate system through a rotation matrix $\mathbf{R}_1$ defined by the angle $\alpha$ (from the global to local system, counterclockwise):
\begin{align}
\boldsymbol{\sigma}_{i'j'}^{\text{ini}} &= \begin{bmatrix}
    \sigma_{x'x'}^{\rm{ini}} & \tau_{x'z'}^{\rm{ini}} \\
    \tau_{z'x'}^{\rm{ini}} & \sigma_{z'z'}^{\rm{ini}}
\end{bmatrix}
=\mathbf{R}_1 \, \boldsymbol{\sigma}_{ij}^{\text{ini}} \, \mathbf{R}_1^{\text{T}}, \label{eq:transforming} 
\end{align}

\noindent where $\mathbf{R}_1 = 
\begin{bmatrix}
\cos \alpha & \sin \alpha \\
- \sin \alpha & \cos \alpha
\end{bmatrix}$ and $\boldsymbol{\sigma}_{ij}^{\rm{ini}}=\begin{bmatrix}
    \sigma_{xx}^{\rm{ini}} & \tau_{xz}^{\rm{ini}} \\
    \tau_{zx}^{\rm{ini}} & \sigma_{zz}^{\rm{ini}}
\end{bmatrix}$. In Eq. \ref{eq:coseismic_static_stress_tensor}, $\Delta\sigma_\mathrm{II}$ represents the shear stress drop associated with the fault rupture which is mode $\mathrm{II}$ failure, and $L$ is the down-dip length of the fault. Once $\Delta \sigma_{\rm{II}}$ is specified, the coseismic static stress tensor $\boldsymbol{\sigma}_{i'j'}^{\text{st}}$ can be calculated in the local coordinate system and subsequently rotated into the global $(x,\,z)$ coordinate system via:
\begin{align}
\boldsymbol{\sigma}_{ij}^{{\rm{st}}} =     
\begin{bmatrix}
        \sigma_{xx}^{\text{st}} & \tau_{xz}^{\text{st}} \\
        \tau_{zx}^{\text{st}} & \sigma_{zz}^{\text{st}}   \end{bmatrix} = \mathbf{R}_1^{\rm{T}}\boldsymbol{\sigma}_{i'j'}^{{\rm{st}}}{\mathbf{R}_1}.
\label{eq:trans2}
\end{align}

\noindent This yields the global stress components $\sigma^{\rm{st}}_{xx}$, $\sigma^{\rm{st}}_{zz}$, $\tau^{\rm{st}}_{xz}$, after fault slip. The analytical solution has been examined via a comparison to numerical simulations (see Appendix \ref{appendix:static}).

\subsection{Coseismic dynamic stress perturbation}
\label{section:coseismic_dynamic_solution}
\begin{figure}
	\centering
	\includegraphics[width=.45\textwidth]{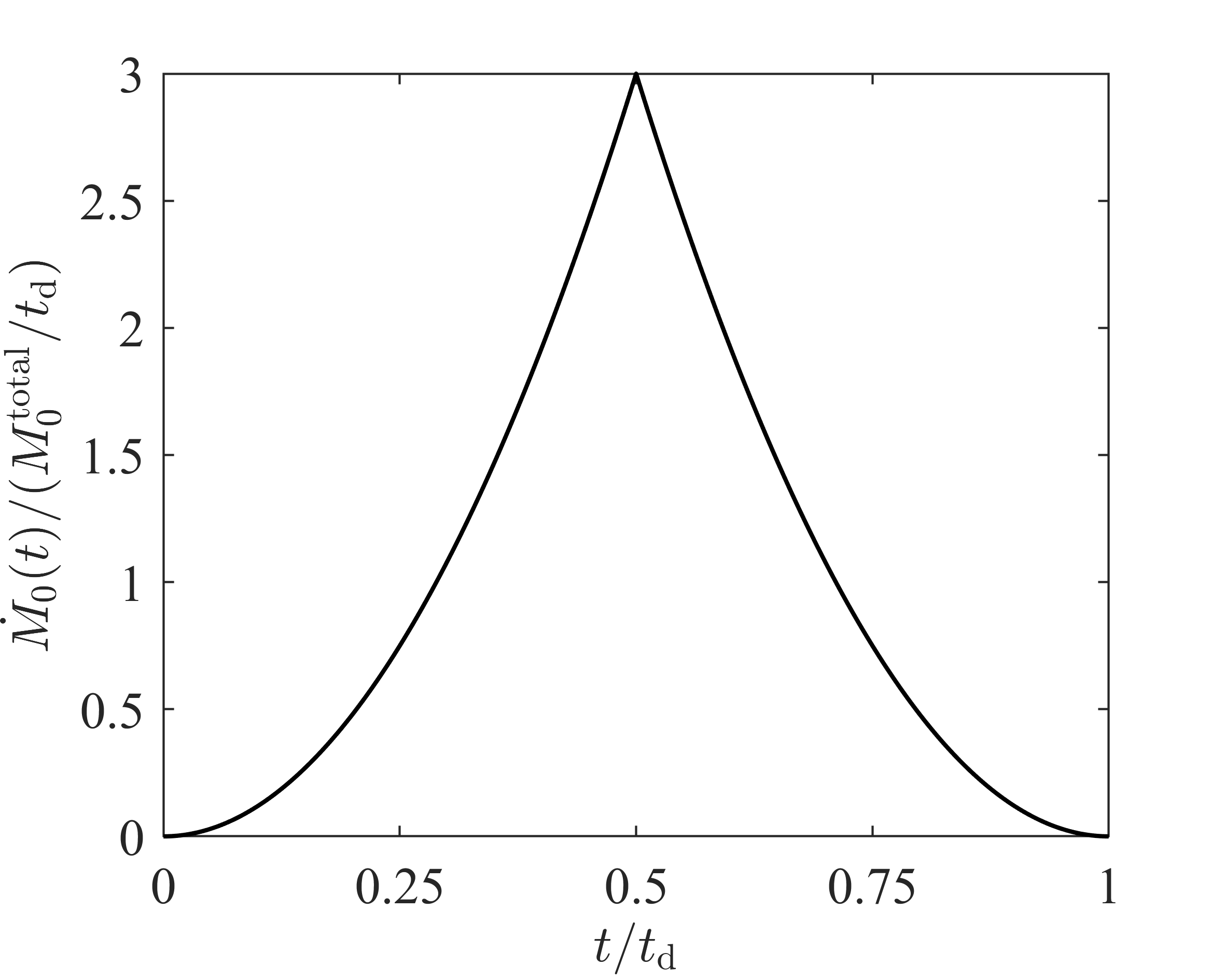}
	\caption{Normalized seismic moment rate as a function of normalized time. The seismic moment rate is normalized by the total seismic moment averaged over the total duration, $M_0^{\rm{total}}/t_{\rm{d}}$, while time is normalized by the total duration $t_{\rm{d}}$. The midpoint in time, $t = t_{\rm{d}}/2$, corresponds to the peak of the normalized seismic moment rate, derived as $C_2$/$C_1$.}
	\label{FIG:SourceTimeFunction}
\end{figure}

The coseismic dynamic displacement field is derived based on seismic source theory \cite{stein2009introduction,lei2021modelling}, yielding:
\begin{subequations} \label{eq:dynamic_displace}
\begin{align}
u_{r'} &= \frac{1}{{4\pi \rho V_{\text{P}}^3r'}}\dot M_0\left( t - \frac{r'}{V_{\text{P}}} \right)\sin 2\theta', \label{eq:dynamic_displace_a} \\
u_{\theta'} &= - \frac{1}{{4\pi \rho V_{\text{S}}^3r'}}\dot M_0\left( t - \frac{r'}{V_{\text{S}}} \right)\cos 2\theta', \label{eq:dynamic_displace_b}
\end{align}
\end{subequations}

\noindent where $u_{r'}$ and $u_{\theta'}$ are the radial and tangential displacement components in the polar coordinate system $(r',\,\theta')$ (see Fig.~\ref{FIG:SchematicModel}b), $\rho$ is rock density, $t$ is time, and $\dot M_0$ is the seismic moment rate, i.e., the time derivative of seismic moment $M_0(t)$. The P- and S-wave velocities, denoted respectively as $V_\text{P}$ and $V_\text{S}$, are given by $V_{\rm{P}} = \sqrt {\frac{{E\left( {1 - \nu } \right)}}{{\rho \left( {1 + \nu } \right)\left( {1 - 2\nu } \right)}}}$ and $V_{\rm{S}} = \sqrt {\frac{E}{{2\rho \left( {1 + \nu } \right)}}}$, where $E$ is Young's modulus and $\nu$ is Poisson's ratio. The slip vector on the fault plane is oriented along the $x'$-direction, such that material above the fault plane $(z'>0)$ is displaced in the positive $x'$ direction. We adopt the source time function $\dot M_0(t)$ proposed by Uchide and Ide (2010) \cite{uchide2010scaling}, which is well-suited for small and medium earthquakes with moment magnitude $M_{\rm{w}}$ of $1.7-4.6$. This function exhibits a symmetric profile, characterized by a growth stage followed by a decline (Fig. \ref{FIG:SourceTimeFunction}). During the growth stage, the cumulative moment ${M_0}(t)$ is given by:
\begin{align}  
    {M_0}\left( t \right) = {C_1}{t^3},
    \label{eq:cumulative_moment} 
\end{align}

\noindent where the constant ${C_1} = 2 \times {10^{17}}\ {\rm{ Nm/}}{{\rm{s}}^3}$. This cumulative moment function is independent of the earthquake magnitude, reflecting the self-similarity of earthquake rupture growth \cite{uchide2010scaling}. The moment rate $\dot M_0(t)$ is further given by:
\begin{align}
\dot{M_0}(t) = 
\begin{cases}
{C_2}{t^2}, & 0 \le t < \dfrac{t_{\mathrm{d}}}{2} \\[6pt]
{C_2}{(t_{\rm{d}}-t)^{2}}, & \dfrac{t_{\mathrm{d}}}{2} \le t < t_{\mathrm{d}}
\end{cases},
\label{eq:source_time}
\end{align}

\noindent where the constant $C_2=6\times{10^{17}}\ {\rm{Nm/{s^{3}}}}$, $t_{\rm{d}}$ is the total duration of the source process including both the growth and decline phases. For a given moment magnitude $M_\mathrm{w}$, the corresponding total seismic moment $M_{0}^{\rm{total}}$ can be derived as\cite{scholz2019mechanics}:
\begin{equation}
{\log _{10}}{M_{0}^{\rm{total}}} = \frac{3}{2}\left( {{M_{\rm{w}}} + 6.07} \right).
\label{eq:moment_magnitude}
\end{equation}
\noindent Here, $M_{0}^{\rm{total}}$ can also be expressed as $M_{0}^{\rm{total}}=\int_0^{{t_{\rm{d}}}} {{{\dot M}_0}\left( t \right)} dt=2{M_0}\left( {{t_{\rm{d}}/2}} \right)$. Based on this formulation, $t_{\rm{d}}$ is determined by:
\begin{align}
    {t_{\rm{d}}} = {C_3}{\left( {M_{0}^{\rm{total}}} \right)^{1/3}},
    \label{eq:duration}
\end{align} 
\noindent where $C_3=2.7\times{10^{-6}}\ {\rm{N}^3{m}^3\rm{s}}$.

By substituting Eqs.~\ref{eq:cumulative_moment}--\ref{eq:duration} into Eq.~\ref{eq:dynamic_displace}, we can compute the time-dependent coseismic dynamic displacement field. The corresponding dynamic stress perturbation $\Delta\boldsymbol{\sigma}_{i'j'}^\mathrm{dy}$ in the polar $(r',\,\theta')$ coordinate system can then be determined based on Hooke’s law assuming linear elasticity \cite{cotton1997dynamic}:
\begin{align}
\Delta \boldsymbol{\sigma}_{i'j'}^{\text{dy}} &=
\begin{bmatrix}
\Delta \sigma_{r'r'}^{\text{dy}} & \Delta \tau_{r'\theta'}^{\text{dy}} \\
\Delta \tau_{\theta'r'}^{\text{dy}} & \Delta \sigma_{\theta'\theta'}^{\text{dy}}
\end{bmatrix}, \label{eq:dynamic_stress} \\
\Delta \sigma_{r'r'}^{\text{dy}} &= (\lambda + 2G) \frac{\partial u_{r'}}{\partial r'} + \frac{\lambda}{r'} \frac{\partial u_{\theta'}}{\partial \theta'} + \frac{\lambda}{r'} u_{r'}, \notag \\
\Delta \sigma_{\theta'\theta'}^{\text{dy}} &= \lambda \frac{\partial u_{r'}}{\partial r'} + \frac{\lambda + 2G}{r'} \frac{\partial u_{\theta'}}{\partial \theta'} + \frac{\lambda + 2G}{r'} u_{r'}, \notag \\
\Delta \tau_{r'\theta'}^{\text{dy}} &= \Delta \tau_{\theta'r'}^{\text{dy}} = G \left[ r' \frac{\partial}{\partial r'}\left( \frac{u_{\theta'}}{r'} \right) + \frac{1}{r'} \frac{\partial u_{r'}}{\partial \theta'} \right], \notag
\end{align}

\noindent where the superscript ‘dy’ indicates the stress disturbance due to coseismic dynamic effects; $\lambda$ and $G$ are the Lamé constants, which are related to Young's modulus $E$ and Poisson's ratio $\nu$ as $\lambda = E \nu/[(1 + \nu)(1 - 2\nu)]$ and $G = E/[2(1 + \nu)]$. The tensor of dynamic stress perturbation $\Delta\boldsymbol{\sigma}_{i'j'}^\mathrm{dy}$ is first transformed from polar $(r',\,\theta')$ coordinates to Cartesian $(x',\,z')$ coordinates through:
\begin{align} \label{eq:dynamic_stress_coordi_trans}
\begin{bmatrix}
\Delta \sigma_{x'x'}^{\mathrm{dy}} & \Delta \tau_{x'z'}^{\mathrm{dy}} \\
\Delta \tau_{z'x'}^{\mathrm{dy}} & \Delta \sigma_{z'z'}^{\mathrm{dy}}
\end{bmatrix}
&= \mathbf{R}_2 \, \Delta \boldsymbol{\sigma}_{i'j'}^{\mathrm{dy}} \, \mathbf{R}_2^{\mathrm{T}},
\end{align}

\noindent where $\mathbf{R}_2 =
\begin{bmatrix}
\cos \theta' & \sin \theta' \\
- \sin \theta' & \cos \theta'
\end{bmatrix}$, and subsequently transformed to the global $(x,\,z)$ coordinate system as:
\begin{align}
    \Delta \boldsymbol{\sigma} _{ij}^{{\rm{dy}}} = \begin{bmatrix}
       {\Delta \sigma _{xx}^{{\rm{dy}}}}&{\Delta \tau _{xz}^{{\rm{dy}}}}\\
       {\Delta \tau _{zx}^{{\rm{dy}}}}&{\Delta \sigma _{zz}^{{\rm{dy}}}} 
    \end{bmatrix} = \mathbf{R}_1^{\rm{T}}\begin{bmatrix}
    {\Delta \sigma _{x'x'}^{{\rm{dy}}}}&{\Delta \tau _{x'z'}^{{\rm{dy}}}}\\
    {\Delta \tau _{z'x'}^{{\rm{dy}}}}&{\Delta \sigma _{z'z'}^{{\rm{dy}}}}
\end{bmatrix}{\mathbf{R}_1}.\label{transform3}
\end{align}

\noindent Therefore, the final form of the dynamic stress field $\boldsymbol{\sigma}_{ij}^\mathrm{dy}$ is given by the superimposition of the dynamic stress disturbance and the initial stress field:
\begin{align}
    \boldsymbol{\sigma}_{ij}^{{\rm{dy}}} = \begin{bmatrix}
        {\sigma _{xx}^{{\rm{dy}}}}&{\tau _{xz}^{{\rm{dy}}}}\\
        {\tau _{zx}^{{\rm{dy}}}}&{\sigma _{zz}^{{\rm{dy}}}}
    \end{bmatrix} = \boldsymbol{\sigma}_{ij}^{{\rm{ini}}} + \Delta \boldsymbol{\sigma}_{ij}^{{\rm{dy}}}. \label{eq:dynamic_stress_V2}
\end{align}

\noindent The above solution of dynamic stress tensor has been examined via a comparison against the numerical simulation results of a point source problem \cite{wang2017numerical} (see Appendix \ref{appendix:dynamic}).

While the seismic source theory provides a robust framework for modeling far-field seismic radiation, it shows significant limitations when being applied to near-field problems, particularly in capturing the effects of a finite fault. To address this issue, we discretize the fault into $N$ segments, with each segment represented by a point source located at its center (Fig.~\ref{FIG:SchematicModel}c) \cite{lei2022reply}. Assuming a uniform slip distribution along the fault, each segment radiates an equal amount of seismic energy. The total seismic moment released by the fault can then be derived as ${M_{0}^{\rm{total}}} = \sum\limits_{i = 1}^N {M_{0,i}^{\rm{total}}}$, where $M_{0,i}^{\rm{total}}$ is the total seismic moment of the $i$th segment. Since the dimension of each segment is much smaller than its distance to the target tunnel, the coseismic response at the target tunnel can be approximated as far-field response. Thus, the single-point source solution, which is well established for far-field analysis, can be used to approximate the coseismic dynamic effects at the target tunnel induced by each segment. Consequently, Eq.~\ref{eq:dynamic_stress_coordi_trans} can be reformulated to incorporate the superposition of dynamic stress contributions from all segments:

\begin{align} 
\begin{aligned}
&\begin{bmatrix}
\Delta \sigma_{x'x'}^{\mathrm{dy}} & \Delta \tau_{x'z'}^{\mathrm{dy}} \\
\Delta \tau_{z'x'}^{\mathrm{dy}} & \Delta \sigma_{z'z'}^{\mathrm{dy}}
\end{bmatrix}=
\begin{bmatrix}
\sum\limits_{k=1}^N \Delta \sigma_{x'_k x'_k}^{\rm dy} & 
\sum\limits_{k=1}^N \Delta \tau_{x'_k z'_k}^{\rm dy} \\
\sum\limits_{k=1}^N \Delta \tau_{z'_k x'_k}^{\rm dy} & 
\sum\limits_{k=1}^N \Delta \sigma_{z'_k z'_k}^{\rm dy}
\end{bmatrix} =
\begin{bmatrix}
\sum\limits_{k=1}^N \mathbf{R}_{2,k} \Delta \sigma_{r'_k r'_k}^{\rm dy} \mathbf{R}_{2,k}^{\rm T} & 
\sum\limits_{k=1}^N \mathbf{R}_{2,k} \Delta \tau_{r'_k \theta'_k}^{\rm dy} \mathbf{R}_{2,k}^{\rm T} \\
\sum\limits_{k=1}^N \mathbf{R}_{2,k} \Delta \tau_{\theta'_k r'_k}^{\rm dy} \mathbf{R}_{2,k}^{\rm T} & 
\sum\limits_{k=1}^N \mathbf{R}_{2,k} \Delta \sigma_{\theta'_k \theta'_k}^{\rm dy} \mathbf{R}_{2,k}^{\rm T}
\end{bmatrix}, \label{eq:dynamic_stress_superposition}
\end{aligned}
\end{align}
\noindent where the rotation matrix $\mathbf{R}_{2,k}$ for the $k_{\rm{th}}$ discretized point source is defined as $\mathbf{R}_{2,k} =
\begin{bmatrix}
\cos \theta'_k & \sin \theta'_k \\
-\sin \theta'_k & \cos \theta'_k
\end{bmatrix}$. This rotation matrix transforms the stress components, originally expressed in the polar coordinate system $(r_k^\prime,\theta_k^\prime)$, centered at the $k_{\rm{th}}$ point source (Fig.~\ref{FIG:SchematicModel}c), into the corresponding stress components in the Cartesian system $(x_k^\prime,z_k^\prime)$ for tensorial superposition. The stress components $\Delta \sigma_{r'_k r'_k}^{\rm dy}$, $\Delta \sigma_{\theta'_k \theta'_k}^{\rm dy}$, and $\Delta \tau_{r'_k \theta'_k}^{\rm dy}$ are obtained from Eqs. \ref{eq:dynamic_displace} and \ref{eq:dynamic_stress} using the polar coordinates $(r_k^\prime,\theta_k^\prime)$.

\subsection{Stress field around a tunnel under coseismic static and dynamic effects}

Consider a circular tunnel of radius $a$ with a free-surface boundary condition, where the radial and shear stresses, $\sigma_{rr}^{\rm{tun}}$ and $\tau_{r\theta}^{\rm{tun}}$, at the tunnel surface are zero. The superscript `tun' denotes the stress state surrounding the tunnel. Taking $\boldsymbol{\sigma}_{ij}^{\rm{st}}$ and $\boldsymbol{\sigma}_{ij}^{\rm{dy}}$ as the far-field stress condition, we can compute the redistributed coseismic stress field around the tunnel in polar coordinates $(r,\,\theta)$ based on the Kirsch solution \cite{jaeger2009fundamentals}:
\begin{subequations}
\label{eq:stress}
\begin{align}
\sigma_{rr}^{\text{tun}} 
&= \frac{\sigma_{xx}^{\mathrm{st/dy}} + \sigma_{zz}^{\mathrm{st/dy}}}{2} 
\left(1 - \frac{a^2}{r^2} \right) + \frac{\sigma_{xx}^{\mathrm{st/dy}} - \sigma_{zz}^{\mathrm{st/dy}}}{2} 
 \left(1 - 4\frac{a^2}{r^2} + 3\frac{a^4}{r^4} \right) \cos{2\theta} + \tau_{xz}^{\mathrm{st/dy}} 
 \left(1 - 4\frac{a^2}{r^2} + 3\frac{a^4}{r^4} \right) \sin{2\theta},\label{eq:stress_a}
 \\
\sigma_{\theta\theta}^{\text{tun}} &= \frac{\sigma_{xx}^{\mathrm{st/dy}} + \sigma_{zz}^{\mathrm{st/dy}}}{2} \left(1 + \frac{a^2}{r^2} \right)  - \frac{\sigma_{xx}^{\rm{st/dy}} - \sigma_{zz}^{\rm{st}}}{2} \left(1 + 3\frac{a^4}{r^4} \right) \cos 2\theta - \tau_{xz}^{\rm{st/dy}}\left( 1+3\frac{a^4}{r^4}\right) \sin{2\theta},\label{eq:stress_b} \\
\tau_{r\theta}^{\text{tun}} &= \tau_{\theta r}^{\text{tun}} = - \frac{\sigma_{xx}^{\rm{st/dy}} - \sigma_{zz}^{\rm{st/dy}}}{2} \left(1 + 2\frac{a^2}{r^2} - 3\frac{a^4}{r^4} \right) \sin 2\theta + \tau_{xz}^{\rm{st/dy}}\left( 1 + 2\frac{a^2}{r^2}-3\frac{a^4}{r^4} \right)\cos{2\theta},\label{eq:stress_c}
\end{align}
\end{subequations}

\noindent where the superscripts 'st' or 'dy' denote the far-field stress boundaries related respectively to coseismic static or dynamic effects, $r$ is the radial distance from the tunnel center that is located at $(x_{0},z_{0})$, and $\theta$ denotes the polar angle measured counterclockwise from the positive direction of the $x$-axis (Fig.~\ref{FIG:SchematicModel}), given by $r = \sqrt {{{\left( {x - {x_0}} \right)}^2} + {{\left( {z - {z_0}} \right)}^2}}$ and $\theta  = {\rm{atan2}}\left( {z - {z_0},x - {x_0}} \right)$.
The maximum and minimum principal stresses around the tunnel, $\sigma_{1}^{\text{tun}}$ and $\sigma_{3}^{\text{tun}}$, are computed as:
\begin{align}
  \sigma_{1,3}^{\text{tun}} &= \frac{\sigma_{rr}^{\text{tun}} + \sigma_{\theta\theta}^{\text{tun}}}{2} \pm \sqrt{\left( \frac{\sigma_{rr}^{\text{tun}} - \sigma_{\theta\theta}^{\text{tun}}}{2} \right)^2 + \left( \tau_{r\theta}^{\text{tun}} \right)^2}.
\label{placeholder}
\end{align}

In this analytical framework, we assume that the coseismic static and dynamic stresses act as background stresses for the tunnel. This assumption is justified by two main considerations. First, the tunnel (with a typical diameter of $\sim10$ m) is generally much smaller in size compared to the relevant seismogenic fault (with a length of a few hundred meters or more). Second, our analysis focuses on scenarios where the seismogenic fault is located at a distance of more than 10 times of the tunnel size away from the tunnel, representing a typical situation of fault-slip rockburst events as documented at many sites \cite{simser2001geotechnical,rehbock2018fault,husen2012induced}. In such cases, the seismic source is in the far-field zone relative to the tunnel.

To facilitate the subsequent analysis, we define a set of stress quantities to represent the initial stress state and its perturbations induced by coseismic static and dynamic effects. The corresponding notations and definitions are summarized in Table \ref{tbl-symbols} and are used consistently throughout this study.

\begin{table}[h]
\caption{Description of stress quantities used to analyze static and dynamic triggering effects.}
\label{tbl-symbols}
\begin{tabularx}{0.95\linewidth}{@{} l >{\raggedright\arraybackslash}X @{}}
\toprule
\textbf{Notation} & \textbf{Definition}\\
\midrule

$\sigma_1^{\rm{ini}}$ & Maximum principal stress under the initial stress state \\

$\sigma_1^{\rm{st}}$,  $\sigma_1^{\rm{dy}}$ & Maximum principal stresses under coseismic static and dynamic disturbances, respectively\\

$\Delta \sigma_1^{\rm{st}}$,  $\Delta \sigma_1^{\rm{dy}}$ & Change in maximum principal stress induced by coseismic static and dynamic disturbances, respectively, defined as $(\sigma_1^{\rm{st}} - \sigma_1^{\rm{ini}})$  and $(\sigma_1^{\rm{dy}}-\sigma_1^{\rm{ini}})$ \\

$\sigma_1^{\rm{tun,ini}}$ & Maximum principal stress surrounding the tunnel under the initial stress state (after excavation but before the fault-slip event)\\

$\sigma_1^{\rm{tun,st}}, \sigma_1^{\rm{tun,dy}}$ &  Maximum principal stresses surrounding the tunnel under coseismic static and dynamic disturbances, respectively\\

$\Delta\sigma_1^{\rm{tun,st}}$, $\Delta\sigma_1^{\rm{tun,dy}}$  & Changes in the maximum principal stress surrounding the tunnel induced by coseismic static and dynamic disturbances, respectively, defined as $(\sigma_1^{\rm{tun,st}}-\sigma_1^{\rm{tun,ini}})$ and  $(\sigma_1^{\rm{tun,dy}}-\sigma_1^{\rm{tun,ini}})$\\

$\sigma_{\theta\theta}^{\rm{max,ini}}$ & Maximum tangential stress along the boundary of the tunnel under the initial far-field stress condition (after excavation but before the fault-slip event)\\

$\sigma_{\theta\theta}^{\rm{max,st}}$, $\sigma_{\theta\theta}^{\rm{max,dy}}$ & Maximum tangential stress along the boundary of the tunnel under coseismic static and dynamic disturbances, respectively\\

$\sigma_{\theta\theta}^{\rm{peak,dy}}$ & Peak dynamic tangential stress, which is the temporal maximum of $\sigma_{\theta\theta}^{\rm{max,dy}}(t)$\\

$\Delta \sigma_{\rm{trigger}}^{\rm{st}}$, $\Delta \sigma_{\rm{trigger}}^{\rm{dy}}$ & Static and dynamic triggering effects on rockbursts, quantified as the changes in the maximum tangential stress, defined by $(\sigma_{\theta\theta}^{\rm{max,st}}-\sigma_{\theta\theta}^{\rm{max,ini}})$ and $(\sigma_{\theta\theta}^{\rm{peak,dy}}-\sigma_{\theta\theta}^{\rm{max,ini}})$, respectively\\

\bottomrule
\end{tabularx}
\end{table}

\section{Demonstration of the analytical framework using synthetic cases}
\label{section:analytical-results}

\subsection{Model setup and parameters}
\label{Section:Fault-slip-description}
We demonstrate the analytical framework using a synthetic study, where a seismogenic fault dips at an angle of $\alpha$ = 45$^\circ$, with a hypocenter depth of $h_0$ = 2,000 m, and a fault length of $L$ = 500 m. The fault is assumed to have a circular geometry, with its strike oriented perpendicular to the $x-z$ plane, and its center located within the plane. The total seismic moment $M_{0}^{\rm{total}}$ that this fault can accommodate is estimated using the following empirical scaling relationship \cite{leonard2014self}:
\begin{align}
    {\log _{10}} {{M_{0}^{\rm{total}}}} = 6.38 + 1.5{\log _{10}} A,
    \label{eq:emprical_leonard}
\end{align}
\noindent where $A$ is the fault area. The corresponding shear stress drop $\Delta \sigma_{\rm{II}}$ is estimated using the following theoretical formulation \cite{kanamori1975theoretical}:
\begin{align}
    \Delta {\sigma _{\mathrm{II}}} = \frac{7}{{16}} \cdot \frac{{{M_{0}^{\rm{total}}}}}{{{(L/2)^3}}}.
\end{align}

Applying these relationships, we obtain a total seismic moment of $M_{0}^{\rm{total}} = 2.09\times10^{14}\ \rm{Nm}$, a moment magnitude of $M_{\mathrm{w}} = 3.48$, and a shear stress drop of $\Delta \sigma_{\rm{II}}=5.84\ \rm{MPa}$. The fault is embedded in the rock mass characterized by a density $\rho$ of 2,600 $\rm{kg/m^3}$, a Young's modulus $E$ of 18.97 GPa, and a Poisson's ratio $\nu$ of 0.19 \cite{gao2021numerical,arzua2013dilation}, within which a tunnel with a radius $a$ of 5 m is placed at different locations across the domain, covering both the near-field and far-field regions of the fault. The vertical stress is assumed to be governed by gravity, while the horizontal stress is defined by a horizontal-to-vertical stress ratio of 0.5 \cite{brown1980underground}. Accordingly, the initial stress tensor is given by:
\begin{align}
\boldsymbol{\sigma}_{ij}^{\rm{ini}}=\begin{bmatrix}
    0.5\rho g(h_{0}-z) & 0 \\
    0 & \rho g(h_{0}-z)
\end{bmatrix}.
\label{eq:initial_stress_condition}
\end{align}

\noindent where $g$ is the gravitational acceleration and $h_0-z$ gives the burial depth (see Fig. \ref{FIG:SchematicModel}). The seismogenic fault is discretized into $N$ = 1,000 segments, with each treated as a point source. The chosen level of discretization is proven sufficient for representing seismic wave radiation during the fault rupture (see Appendix \ref{appendix:determination-N}). Using the analytical framework described in Section \ref{section:analytical_model}, both the coseismic static and dynamic stress changes in the surrounding rock mass induced by fault slip can be computed.

\subsection{Coseismic static stress redistribution}
\begin{figure*}
	\centering
	\includegraphics[width=1.0\textwidth]{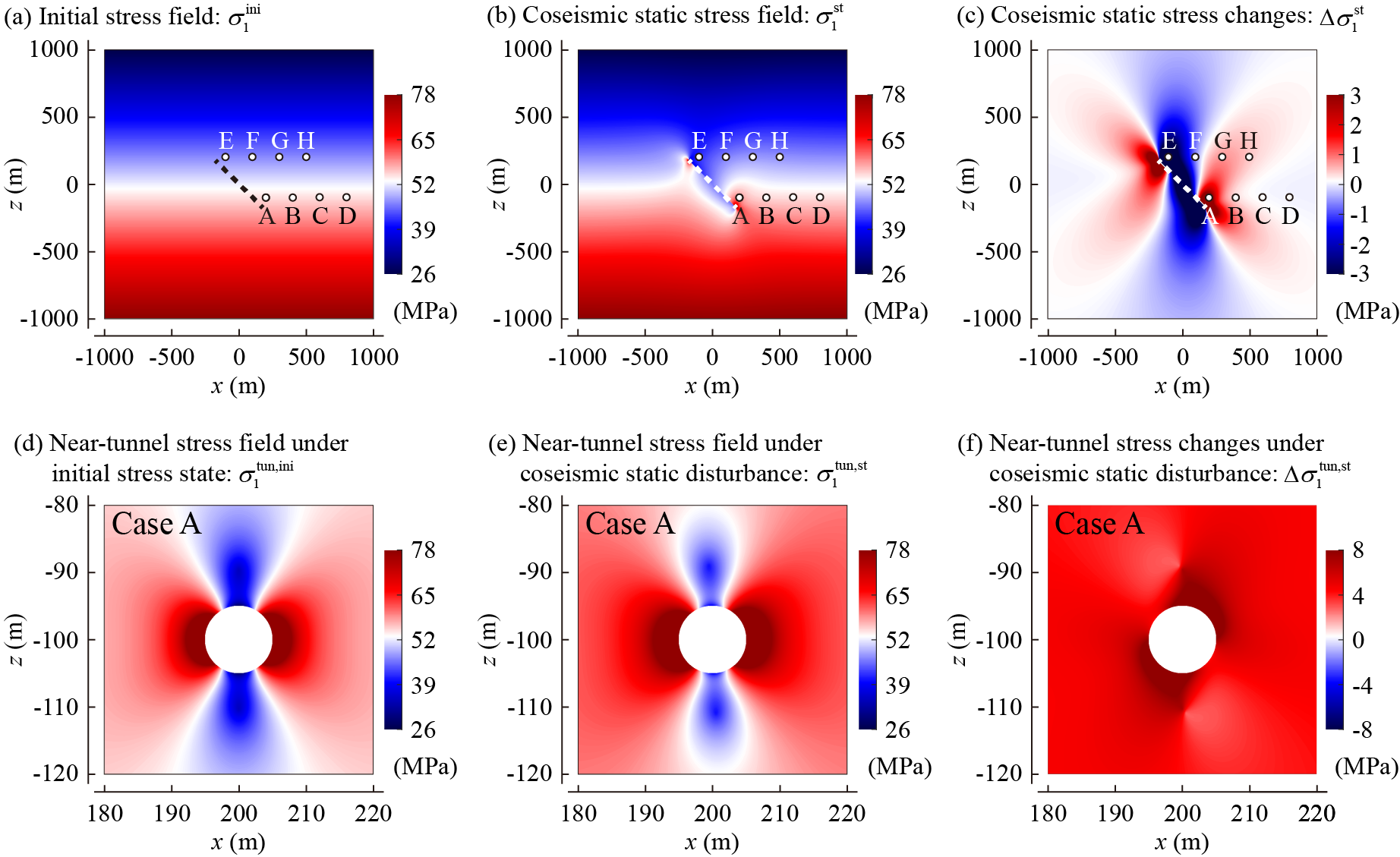} 
	\caption{Maximum principal stress fields under initial and coseismic static conditions. (a-b) Initial  $(\sigma_1^{\rm{ini}})$ and coseismic static $(\sigma_1^{\rm{st}})$ stress fields in the large-scale region, where $x \in [-1000,1000]$, $z\in[-1000,1000]$. (d-e) Initial $(\sigma_1^{\rm{tun,ini}})$ and coseismic static $(\sigma_1^{\rm{tun,st}})$ stress fields in the small-scale region surrounding the tunnel in Case A, where $x \in [180,220]$, $z \in [-120,-80]$. Panels (c) and (f) show the coseismic static stress changes, defined as $\Delta \sigma_1^{\rm{st}}=\sigma_1^{\rm{st}}-\sigma_1^{\rm{ini}}$ and $\Delta \sigma_1^{\rm{tun,st}}=\sigma_1^{\rm{tun,st}}-\sigma_1^{\rm{tun,ini}}$, in the corresponding regions. In panels (a-c), the dashed line represents the fault, and the circles labeled A-H indicate the positions of different tunnel cases, with their radii enlarged to five times the actual tunnel radius for clarity.}
    \label{FIG:Static_Stress_Changes_in_a_Case}
\end{figure*}

Figs. \ref{FIG:Static_Stress_Changes_in_a_Case}a and \ref{FIG:Static_Stress_Changes_in_a_Case}b present the stress fields before ($\sigma_1^{\rm{ini}}$) and after ($\sigma_1^{\rm{st}}$) the coseismic static disturbance. The maximum principal stress is used to characterize the stress state due to its strong relevance to rockburst initiation \cite{gong2023strength,he2012studies,zhou2018evaluation,he2023review}. The coseismic static stress redistribution is characterized by stress concentration near the fault tips and stress reduction in the laterals of the fault. The resulting stress change, defined as $\Delta \sigma_1^{\rm{st}}=\sigma_1^{\rm{st}}-\sigma_1^{\rm{ini}}$ (see Table \ref{tbl-symbols} for detailed definitions of $\sigma_1^{\rm{ini}}$, $\sigma_1^{\rm{st}}$, and $\Delta \sigma_1^{\rm{st}}$), forms lobe-shaped zones surrounding the fault (Fig. \ref{FIG:Static_Stress_Changes_in_a_Case}c), consisting of four compressive lobes with increased stress (highlighted in red) and two dilational lobes with decreased stress (highlighted in blue). We then focus on a specific tunnel, marked as Case A in Fig. \ref{FIG:Static_Stress_Changes_in_a_Case}c. Under the initial background stress condition and after excavation, the stress field around the tunnel ($\sigma_1^{\rm{tun,ini}}$) is computed as shown in Fig. \ref{FIG:Static_Stress_Changes_in_a_Case}d. It reveals substantial compressive loading in the horizontal direction (aligned with the initial minimum principal stress) and tensile unloading in the vertical direction (aligned with the initial maximum principal stress). Following the coseismic static stress perturbation, the compressive zones around the tunnel expand, whereas the tensile zones are suppressed, as shown by $\sigma_1^{\rm{tun,st}}$ in Fig. \ref{FIG:Static_Stress_Changes_in_a_Case}e. This behavior arises from the additional stress changes induced by the coseismic static effects, defined as $\Delta \sigma_1^{\rm{tun,st}}=\sigma_1^{\rm{tun,st}}-\sigma_1^{\rm{tun,ini}}$ (Fig. \ref{FIG:Static_Stress_Changes_in_a_Case}f) (see Table \ref{tbl-symbols} for detailed definitions of $\sigma_1^{\rm{tun,ini}}$, $\sigma_1^{\rm{tun,st}}$, and $\Delta \sigma_1^{\rm{tun,st}}$). These changes exhibit compressive characteristics, particularly from the tunnel crown to the right sidewall, and from the invert to the left sidewall.

\begin{figure*}
	\centering
	\includegraphics[width=1.0\textwidth]{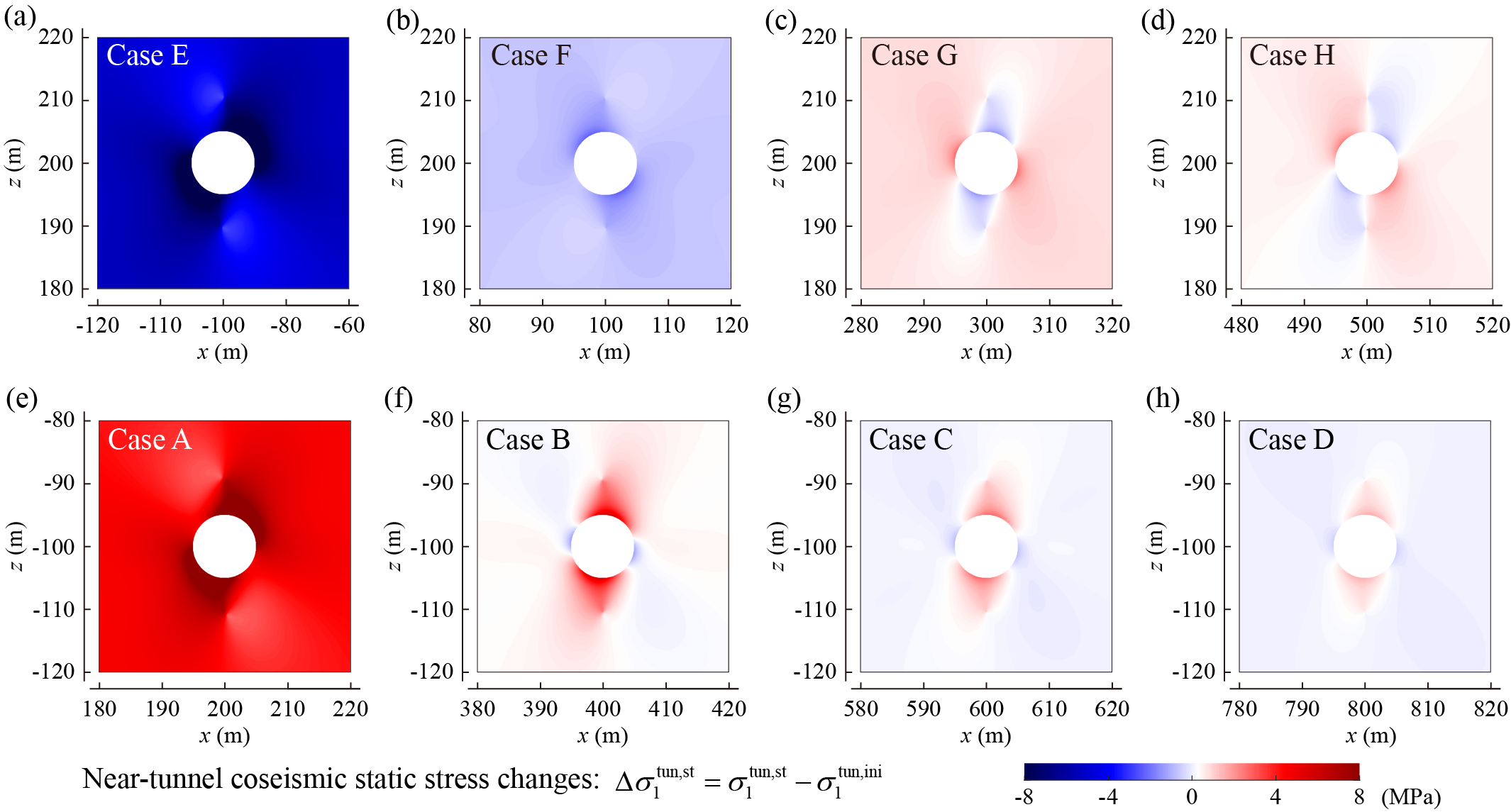}
	\caption{Coseismic static stress changes surrounding the tunnel, defined as $\Delta \sigma_1^{\rm{tun,st}}=\sigma_1^{\rm{tun,st}}-\sigma_1^{\rm{tun,ini}}$, for Cases A-H. The corresponding tunnel positions are indicated in Figs. \ref{FIG:Static_Stress_Changes_in_a_Case}a-c.}
    \label{FIG:Static_Stress_Changes_in_different_Cases}
\end{figure*}

We then investigate the coseismic static stress changes near the tunnel, denoted as $\Delta \sigma_1^{\rm{tun,st}}$, across different tunnel locations, as shown in Fig. \ref{FIG:Static_Stress_Changes_in_different_Cases}. Cases A-D (Figs. \ref{FIG:Static_Stress_Changes_in_different_Cases}e-h), all situated at a burial depth of $h_{0}-z=2100\ \rm{m}$, exhibit additional stress concentrations at the crown and invert, with the intensity of these concentrations increasing as the tunnel is positioned closer to the seismogenic fault. In contrast, Cases E-H (Figs. \ref{FIG:Static_Stress_Changes_in_different_Cases}a-d), located at a burial depth of $h_{0}-z=1800\ \rm{m}$, demonstrate distinct responses. Specifically, Cases E and F, which are positioned near the fault within the coseismic dilational lobes (see Fig. \ref{FIG:Static_Stress_Changes_in_a_Case}c), experience significant coseismic stress unloading. Meanwhile, Cases G and H, though positioned farther from the fault, fall within the compressive lobes and exhibit additional stress concentrations near the tunnel sidewalls. These observations suggest that the coseismic response of a tunnel is influenced not only by its distance from the fault, but also by its spatial position with respect to the fault slip-induced stress redistribution pattern. Thus, this location-dependent effect determines whether the tunnel experiences coseismic stress concentration or release, as well as the spatial distribution of these effects around the tunnel.

\subsection{Coseismic dynamic stress perturbations}
\begin{figure*}
	\centering
	\includegraphics[width=1.0\textwidth]{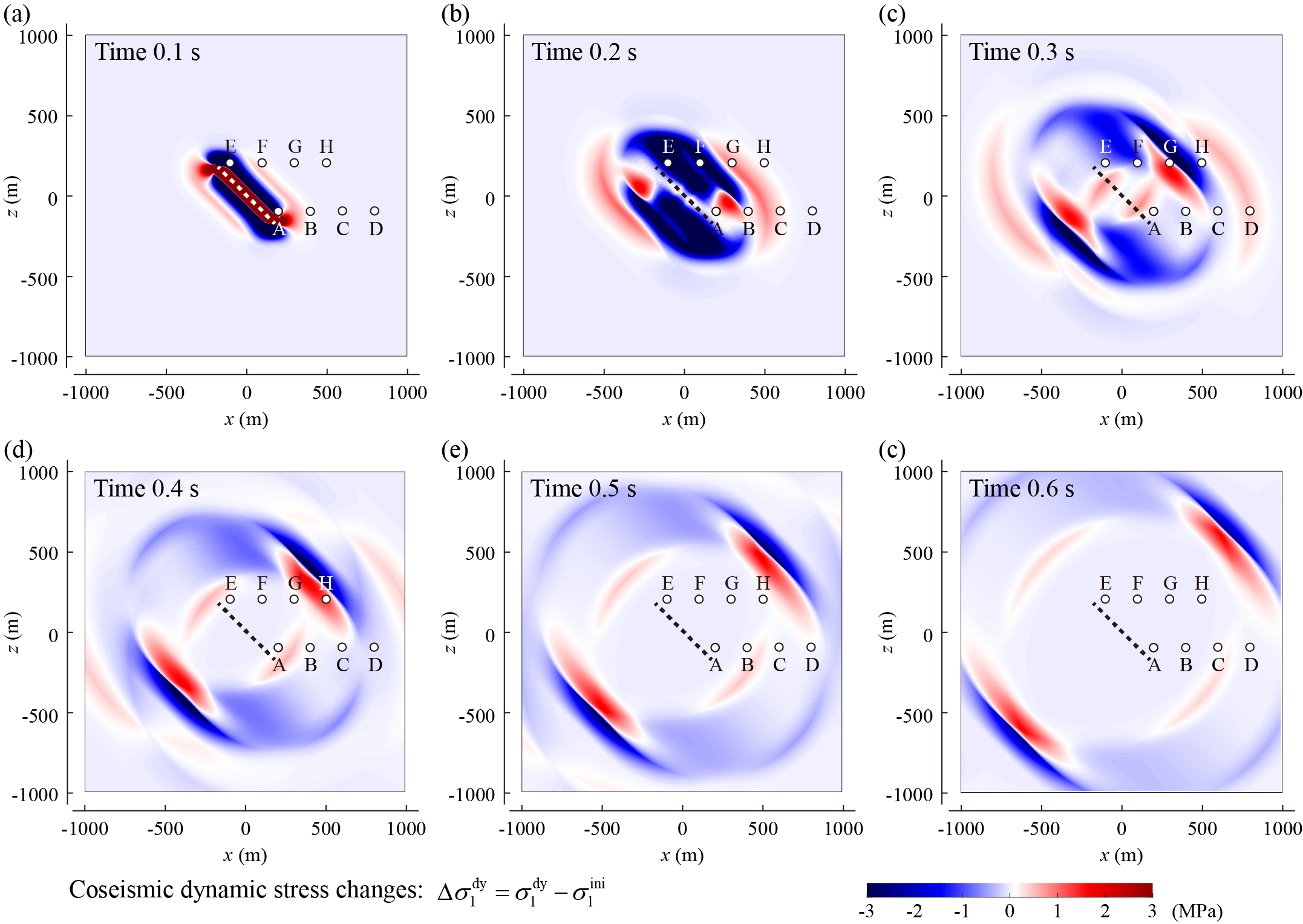}
	\caption{Coseismic dynamic stress changes, defined as $\Delta \sigma_1^{\rm{dy}}=\sigma_1^{\rm{dy}}-\sigma_1^{\rm{ini}}$, in the large-scale region at time instances of 0.1-0.6 s. The dashed line represents the fault, and the circles labeled A-H indicate the positions of different tunnel cases, with their radii enlarged to five times the actual tunnel radius for clarity.}
    \label{FIG:Dynamic_Stress_Changes_in_large_scale}
\end{figure*}

\begin{figure*}
	\centering
	\includegraphics[width=1.0\textwidth]{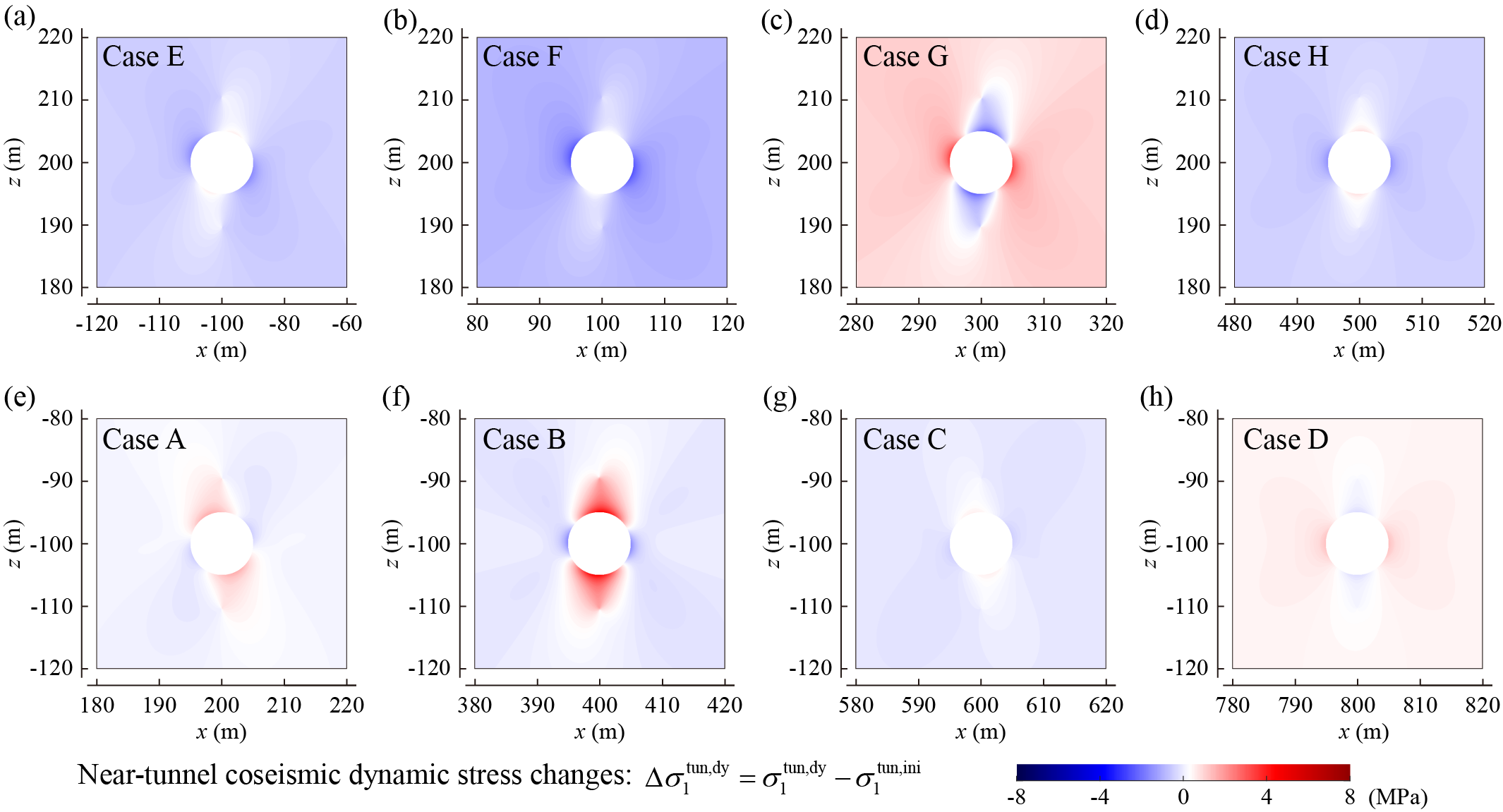}
	\caption{Coseismic dynamic stress changes surrounding the tunnel, defined as $\Delta \sigma_1^{\rm{tun,dy}} =  \sigma_1^{\rm{tun,dy}}-\sigma_1^{\rm{tun,ini}}$, for different cases at the time instance of 0.3 s. The corresponding tunnel positions are indicated in Fig. \ref{FIG:Dynamic_Stress_Changes_in_large_scale}c.}
    \label{FIG:Dynamic_Stress_Changes_in_different_Cases}
\end{figure*}

Fig. \ref{FIG:Dynamic_Stress_Changes_in_large_scale} illustrates the temporal evolution of coseismic dynamic stress waves, represented by $\Delta \sigma_1^{\rm{dy}}$ (see Table \ref{tbl-symbols} for definition). These waves, consisting of both compressive and dilational components, propagate in an antisymmetric pattern with respect to the seismogenic fault. Specifically, wave amplitudes are stronger in the direction normal to the fault plane and are significantly attenuated in the down-dip direction of the fault. As these dynamic waves interact with a tunnel, they induce transient stress concentrations or stress releases in the surrounding rock mass. Fig. \ref{FIG:Dynamic_Stress_Changes_in_different_Cases} presents the stress field evolution, defined as $\Delta \sigma_1^{\rm{}tun,dy}$ (see Table \ref{tbl-symbols} for definition), around various tunnel cases at a time instance of 0.3 s. Case G exhibits pronounced stress concentrations at the sidewalls, while Case B shows strong stress concentrations at the crown and invert, both resulting from the influence of compressive waves associated with S-wave propagation perpendicular to the fault (refer to Fig. \ref{FIG:Dynamic_Stress_Changes_in_large_scale}c). Case A experiences slight stress concentration due to S-waves propagating along the fault's down-dip direction, whereas Case D shows stress concentrations at the sidewall primarily driven by compressive waves. The remaining cases, C, E, F, H, are predominantly subjected to stress release, as they are located within the zones where dilational waves dominate at this specific time.

\subsection{Static versus dynamic triggering effects on rockbursts}
\begin{figure*}
	\centering
	\includegraphics[width=1.0\textwidth]{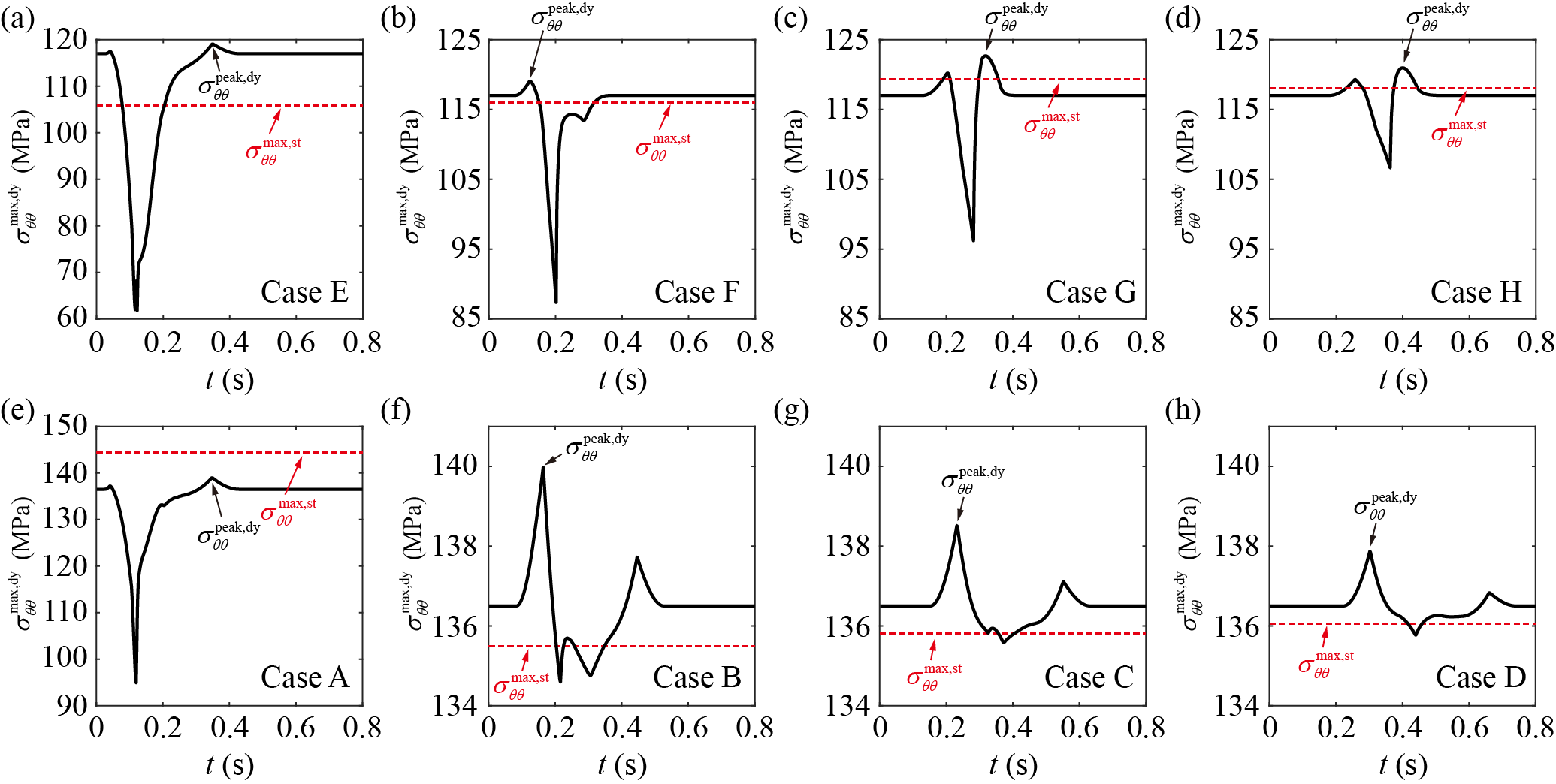}
	\caption{Temporal evolution of the maximum tangential stress on the boundary of the tunnel under coseismic dynamic disturbance, defined as $\sigma_{\theta\theta}^{\rm{max,dy}}$, for different cases. The corresponding tunnel positions are shown in Fig. \ref{FIG:Dynamic_Stress_Changes_in_large_scale}. The red dashed line represents the maximum tangential stress on the boundary of the tunnel under coseismic static disturbance, defined as $\sigma_{\theta\theta}^{\rm{max,st}}$, for each case.}
    \label{FIG:Dynamic_Curve_in_different_Cases}
\end{figure*}

Rockbursts commonly occur when the maximum tangential stress ($\sigma_{\theta\theta}^{\rm{max}}$) acting along the tunnel boundary approaches or exceeds the uniaxial compressive strength ($\sigma_{\rm{c}}$) of the surrounding rock mass \cite{dowding1986potential,li2021principles,wang2021numerical,he2023review}. To evaluate the rockburst potential or susceptibility, we track the maximum tangential stress around the tunnel, including the static and dynamic components  $\sigma_{\theta\theta}^{\mathrm{max,st}}$ and $\sigma_{\theta\theta}^{\mathrm{max,dy}}$ respectively under coseismic static and dynamic triggering effects. The results for Cases A-H are shown in Fig. \ref{FIG:Dynamic_Curve_in_different_Cases}, where the dynamic maximum tangential stress at $t=0$ s corresponds to the initial maximum tangential stress $\sigma_{\theta\theta}^{\mathrm{max,ini}}$ along the tunnel boundary (after excavation but before the fault-slip event). The analysis reveals that dilational dynamic waves can generate substantial unloading of the maximum tangential stress, particularly in Cases A, E and F (Figs. \ref{FIG:Dynamic_Curve_in_different_Cases}e, a, and b), which are located close to the seismogenic fault. In Cases B-D (Figs. \ref{FIG:Dynamic_Curve_in_different_Cases}f-h), two distinct peaks in $\sigma _{\theta \theta }^{{\rm{max,dy}}}$ are observed, corresponding to the arrival of P-waves and S-waves that lead to compressive stress changes, respectively. For Cases G and H (Figs. \ref{FIG:Dynamic_Curve_in_different_Cases}c-d), the stress evolution reflects sequential interactions with P-waves and S-waves, leading to clearly identifiable variations in $\sigma _{\theta \theta }^{{\rm{max,dy}}}$. The peak value of the dynamic maximum tangential stress, denoted by $\sigma _{\theta \theta }^{{\rm{peak,dy}}}$, is extracted by identifying the temporal maximum of $\sigma _{\theta \theta }^{{\rm{max,dy}}}(t)$. A comparison of $\sigma _{\theta \theta }^{{\rm{peak,dy}}}$ and $\sigma _{\theta \theta }^{{\rm{max,st}}}$ (both annotated in Fig. \ref{FIG:Dynamic_Curve_in_different_Cases}) shows that, among all eight cases, only Case A exhibits stress enhancement predominantly due to static triggering $(\rm{i.e.,\ }\sigma _{\theta \theta }^{{\rm{max,st}}}>\sigma _{\theta \theta }^{{\rm{peak,dy}}})$. In contrast, the remaining cases are dominated by dynamic triggering $(\rm{i.e.,\ }\sigma _{\theta \theta }^{{\rm{max,st}}}<\sigma _{\theta \theta }^{{\rm{peak,dy}}})$. Notably, for Cases B-F, the coseismic static maximum tangential stress $\sigma _{\theta \theta }^{{\rm{max,st}}}$ is even lower than the initial stress $\sigma _{\theta \theta }^{{\rm{max,ini}}}$. This reduction arises because the initial stress field produces stress concentration at the tunnel sidewalls (see Fig. \ref{FIG:Static_Stress_Changes_in_a_Case}d), whereas the coseismic static disturbance causes unloading $(\Delta \sigma_1^{\rm{tun,st}}<0)$ at the tunnel sidewalls in Cases B-F (Figs. \ref{FIG:Static_Stress_Changes_in_different_Cases}a-b,f-h).

\begin{figure*}
	\centering
	\includegraphics[width=1.0\textwidth]{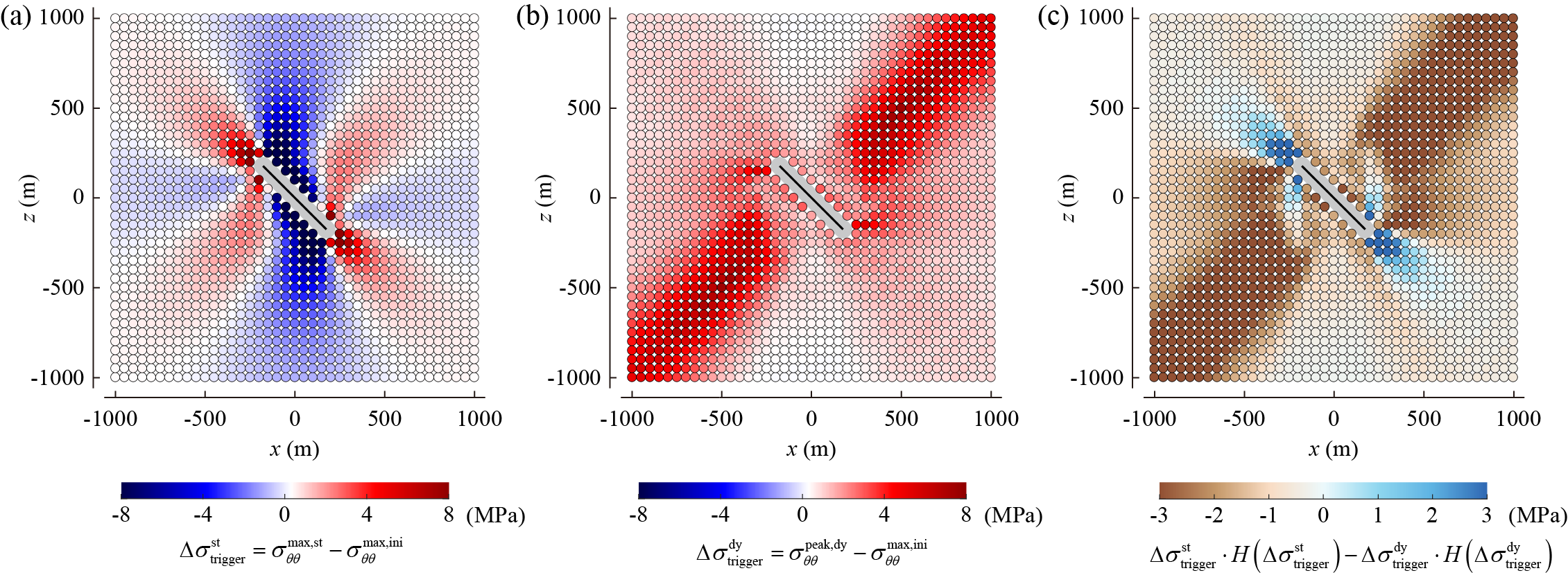}
	\caption{(a) Static triggering effect, (b) dynamic triggering effect, and (c) the competing dominance of static versus dynamic effects on rockburst occurrence in tunnels at various locations. Each circle represents a tunnel case, with its center indicating the tunnel location and its radius enlarged to be five times of the actual tunnel radius for visualization. The contour plots in (a) and (b) indicate the coseismic maximum tangential stress changes, defined as $\Delta \sigma_{\rm{trigger}}^{\rm{st}}$ and $\Delta \sigma_{\rm{trigger}}^{\rm{dy}}$, whereas that in (c) indicates which effect, static or dynamic, is dominant in triggering rockbursts. All tunnel cases are situated outside the gray-shaded region, such that the seismic source far from the tunnel boundary (with distance greater than 10 times the tunnel diameter).}
	\label{FIG:Triggering-role}
\end{figure*}

To evaluate the respective roles of static and dynamic triggering on rockburst initiation in tunnels, we define the static and dynamic triggering stress increments, $\Delta \sigma_{\rm{trigger}}^{\rm{st}}$ and $\Delta \sigma_{\rm{trigger}}^{\rm{dy}}$, given as:
\begin{subequations}
\label{eq:triggering_stresses}
\begin{align}
\Delta \sigma _{{\rm{trigger}}}^{{\rm{st}}} &= \sigma _{\theta \theta }^{{\rm{max,st}}} - \sigma _{\theta \theta }^{{\rm{max,ini}}}, \label{eq:triggering_stresses_a} \\
\Delta \sigma _{{\rm{trigger}}}^{{\rm{dy}}} &= \sigma _{\theta \theta }^{{\rm{peak,dy}}} - \sigma _{\theta \theta }^{{\rm{max,ini}}}. \label{eq:triggering_stresses_b}
\end{align}
\end{subequations}
These two quantities represent the changes in the maximum tangential stress along the tunnel boundary induced by coseismic static and dynamic effects, respectively. We employ the following metric to assess the competition of static versus dynamic triggering:
\begin{align}
    \Delta \sigma _{{\rm{trigger}}}^{{\rm{st}}}H\left( {\Delta \sigma _{{\rm{trigger}}}^{{\rm{st}}}} \right) - \Delta \sigma _{{\rm{trigger}}}^{{\rm{dy}}}H\left( {\Delta \sigma _{{\rm{trigger}}}^{{\rm{dy}}}} \right),
\end{align}
where positive values indicate that static triggering dominates, and negative values indicate that dynamic triggering dominates; $H\left(\cdot\right)$ is the Heaviside step function, defined as $H(x)$ is 0 for $\ x \le 0$ and 1 otherwise. To systematically evaluate the spatial variability of triggering responses, we discretize the space into a 41$\times$41 grid of possible tunnel locations with uniform spacing of 50 m in both horizontal and vertical directions. Each case is modeled with a tunnel having a diameter of $a = 10\ \rm{m}$. The results, illustrating the static triggering effects, dynamic triggering effects, and the competition between the two mechanisms across the study domain, are presented in Fig. \ref{FIG:Triggering-role}.

The static triggering effects give rise to distinct response regions, including four high-risk lobes characterized by positive values of $\Delta \sigma_{\rm{trigger}}^{\rm{st}}$, and four low-risk zones, two aligned horizontally and two vertically, where $\Delta \sigma_{\rm{trigger}}^{\rm{st}}$ is negative. These regions are antisymmetrically distributed with respect to the fault, mirroring the pattern of coseismic static stress changes shown in Fig. \ref{FIG:Static_Stress_Changes_in_a_Case}c. Notably, elevated rockburst risk associated with static triggering is observed near the fault tips along the down-dip direction. In contrast, the dynamic triggering effects primarily intensify rockburst potential along the direction perpendicular to the fault, as well as in local zones near the fault tips. However, within the near-field region along the fault-normal direction, particularly within a distance approximately equal to half of the fault length, the dynamic triggering effect is relatively suppressed. The reduced dynamic triggering potential in this region is attributed to the dominance of dilational seismic waves (see Figs. \ref{FIG:Dynamic_Stress_Changes_in_large_scale}b-c), which induce coseismic stress release at the tunnel sidewalls (Figs. \ref{FIG:Dynamic_Stress_Changes_in_different_Cases}a-b) that have experienced excavation-induced stress concentrations. Fig. \ref{FIG:Triggering-role}c reveals an intense competition between the static and dynamic triggering effects near the fault. Static triggering dominates rockburst potential in tunnels located along the fault's down-tip direction, whereas dynamic triggering dominates along the fault-normal direction. Furthermore, dynamic effects exhibit a broader spatial reach than static effects, suggesting a greater capacity to trigger rockbursts at larger distances from the fault.

\begin{figure*}
	\centering
	\includegraphics[width=1.0\textwidth]{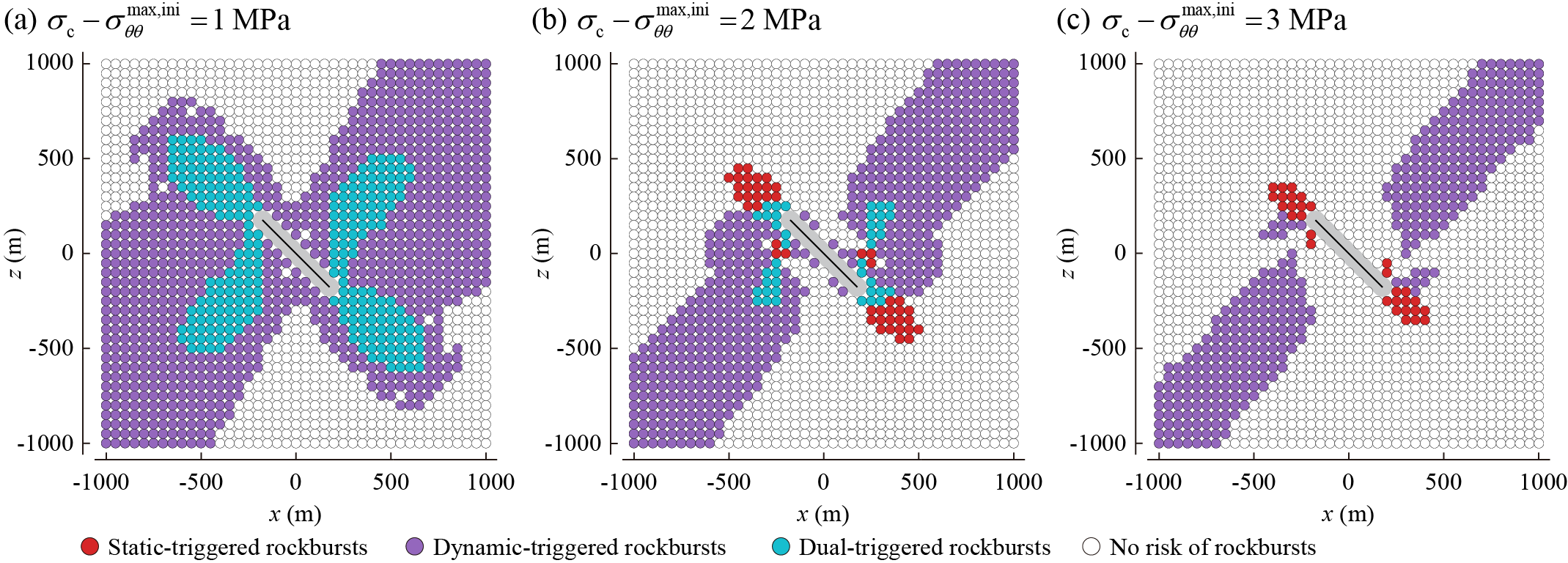}
	\caption{Predicted rockburst occurrences and their associated triggering mechanisms under different stress margins, defined as the difference between the uniaxial compressive strength of the surrounding rock mass $\sigma_{\rm{c}}$ and the initial maximum tangential stress along the tunnel boundary $\sigma_{\theta\theta}^{\rm{max,ini}}$.}
	\label{FIG:Dominant-mechanisms}
\end{figure*}

Based on the above analysis, fault-slip rockburst potential can be evaluated based on the relationships among the initial maximum tangential stress  $\sigma_{\theta\theta}^{\rm{max,ini}}$, the triggering stress increments $\Delta \sigma_{\rm{trigger}}^{\rm{st}}$ and $\Delta \sigma_{\rm{trigger}}^{\rm{dy}}$ (see Table \ref{tbl-symbols} for definitions), and the rock mass strength $\sigma_{\rm{c}}$. Different triggering regimes may be defined as follows:
\begin{itemize}
  \item Static triggering regime: 
  \(
  \Delta \sigma_{\mathrm{trigger}}^{\mathrm{st}} \ge \sigma_{\mathrm{c}} - \sigma_{\theta\theta}^{\mathrm{max,ini}},
  \)
  \item Dynamic triggering regime: 
  \(
  \Delta \sigma_{\mathrm{trigger}}^{\mathrm{dy}} \ge \sigma_{\mathrm{c}} - \sigma_{\theta\theta}^{\mathrm{max,ini}},
  \)
  \item Dual triggering regime:  
  \(
  \min\left\{ \Delta \sigma_{\mathrm{trigger}}^{\mathrm{st}}, \Delta \sigma_{\mathrm{trigger}}^{\mathrm{dy}} \right\} \ge \sigma_{\mathrm{c}} - \sigma_{\theta\theta}^{\mathrm{max,ini}},
  \)
  \item No triggering:  
  \(
  \max\left\{ \Delta \sigma_{\mathrm{trigger}}^{\mathrm{st}}, \Delta \sigma_{\mathrm{trigger}}^{\mathrm{dy}} \right\} < \sigma_{\mathrm{c}} - \sigma_{\theta\theta}^{\mathrm{max,ini}},
  \)
\end{itemize}

\noindent where $\sigma_{\mathrm{c}} - \sigma_{\theta\theta}^{\mathrm{max,ini}}$ defines the stress margin, indicating how close the rock mass is to failure prior to the fault-slip event. We investigate three scenarios with stress margins of 1 MPa, 2 MPa, and 3 MPa (see Fig. \ref{FIG:Dominant-mechanisms}). The results show that a smaller stress margin leads to a larger spatial extent of zones susceptible to rockburst. Under the 1 MPa margin, four-lobed regions emerge where rockbursts may occur under hybrid triggering, i.e., both static and dynamic effects are sufficient to trigger failure. Additionally, in both near-field and far-field regions of the fault, there are areas where dynamic triggering alone could induce rockbursts. At a stress margin of 2 MPa, both dual-triggering and dynamic triggering regions shrink, while some new static triggering zones appear, particularly along the down-dip direction of the fault. When the margin increases to 3 MPa, dual triggering zones disappear entirely. Static triggering-dominated zones are then limited to small areas near the fault tips along the down-dip direction, whereas dynamic triggering-dominated zones remain relatively broadly distributed along the fault-normal direction, except for the immediate vicinity of the fault, where dynamic triggering effects are notably diminished.

\subsection{Influence of model parameters on triggering mechanism}
\begin{figure*}
	\centering
	\includegraphics[width=1.0\textwidth]{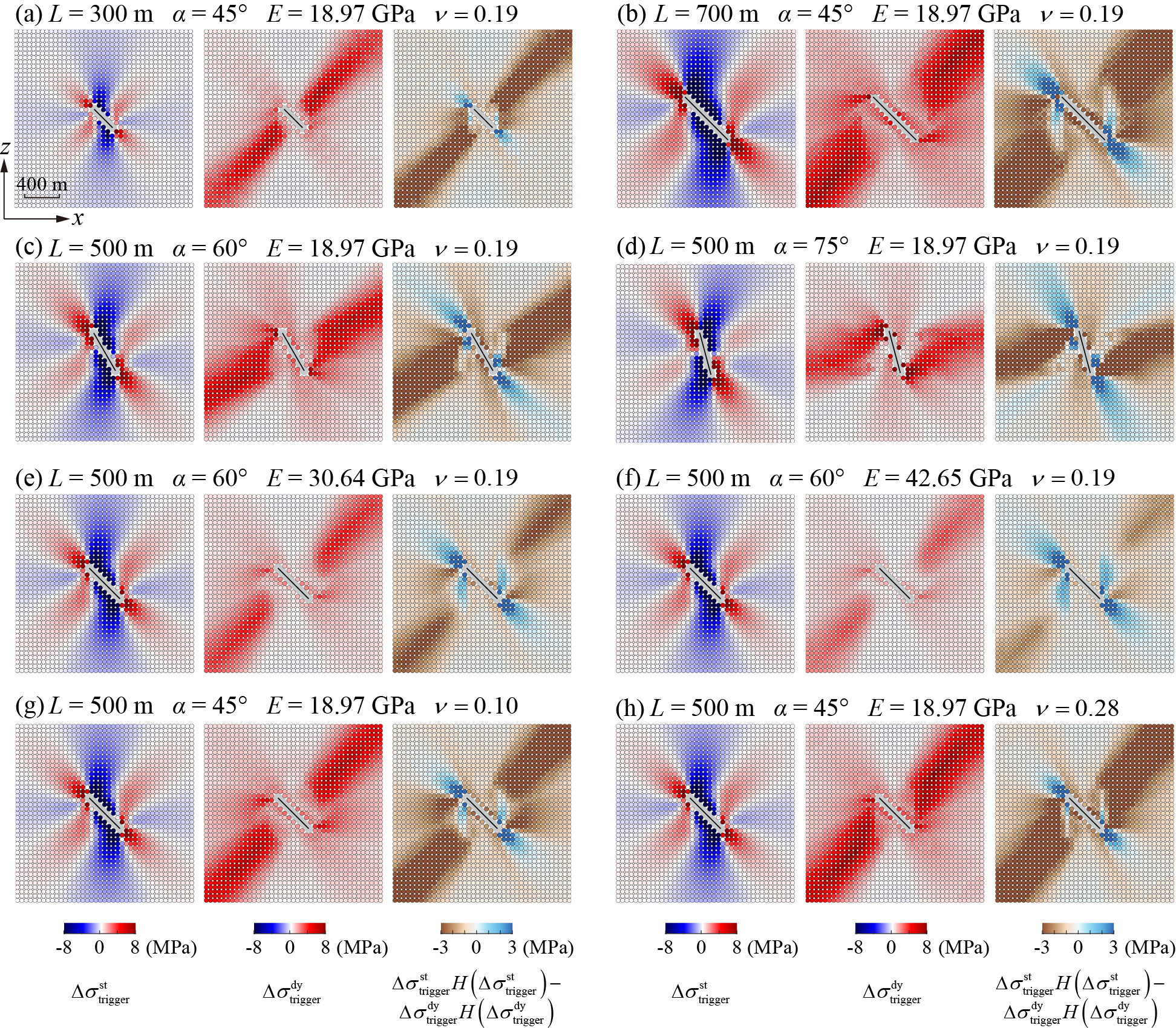}
	\caption{Influence of key parameters on static $(\sigma_{\rm{trigger}}^{\rm{st}})$ and dynamic $(\sigma_{\rm{trigger}}^{\rm{dy}})$ triggering effects, as well as their relative dominance. Panels show the effects of (a-b) fault length $L$, (c-d) fault dip angle $\alpha$, (e-f) Young's modulus $E$, and (g-h) Poisson's ratio $\nu$ of the surrounding rock mass. The ranges of mechanical parameters are informed by strength tests on various types of granite, as reported in Arzúa and Alejano (2013) \cite{arzua2013dilation}.}
	\label{FIG:Parameter_Sensivity}
\end{figure*}

We examine the influence of key parameters of fault geometry and rock materials on static and dynamic triggering mechanisms, as illustrated in Fig. \ref{FIG:Parameter_Sensivity}. Among these factors, fault length plays a critical role in controlling the spatial extent of regions affected by static and/or dynamic triggering (Fig. \ref{FIG:Parameter_Sensivity}a, b). Two representative fault lengths are analyzed: $L=\rm{300\ m}$ and $L=\rm{700\ m}$, corresponding to seismic events with moment magnitudes of $M_{\rm{w}}=3.03$ and $M_{\rm{w}}=3.77$, respectively. In the smaller fault case $(L=\rm{300\ m})$, the rock mass response in the near-field region of the fault driven by static triggering is relatively limited, whereas dynamic effects remain capable of propagating into the far-field region, particularly in the fault-normal direction. In contrast, for the larger fault $(L=\rm{700\ m})$, the spatial extent of static triggering-dominated zones expands significantly. Although dynamic effects cover a broader range in the far-field region, their influence within the near-field region is comparatively weak. The fault dip angle also plays an important role in shaping the orientation of triggering patterns (see Fig. \ref{FIG:Parameter_Sensivity}c, d). Specifically, when the fault has a steeper dip angle (e.g., $\alpha = 75^{\circ}$), dynamic triggering effects become more pronounced near the fault tips. In addition, the mechanical properties of the surrounding rock mass, particularly Young modulus $E$ and Poisson's ratio $\nu$, influence the spatial pattern of dynamic triggering. A higher $E$ and a lower $\nu$ lead to more rapid decay of dynamic triggering, with $E$ exerting a stronger influence than $\nu$. In contrast, static triggering effects are relatively insensitive to these material parameters.

\section{Application to a historical fault-slip rockburst event at the Gotthard Base Tunnel}
\label{section:application}
\begin{figure*}
	\centering
	\includegraphics[width=1.0\textwidth]{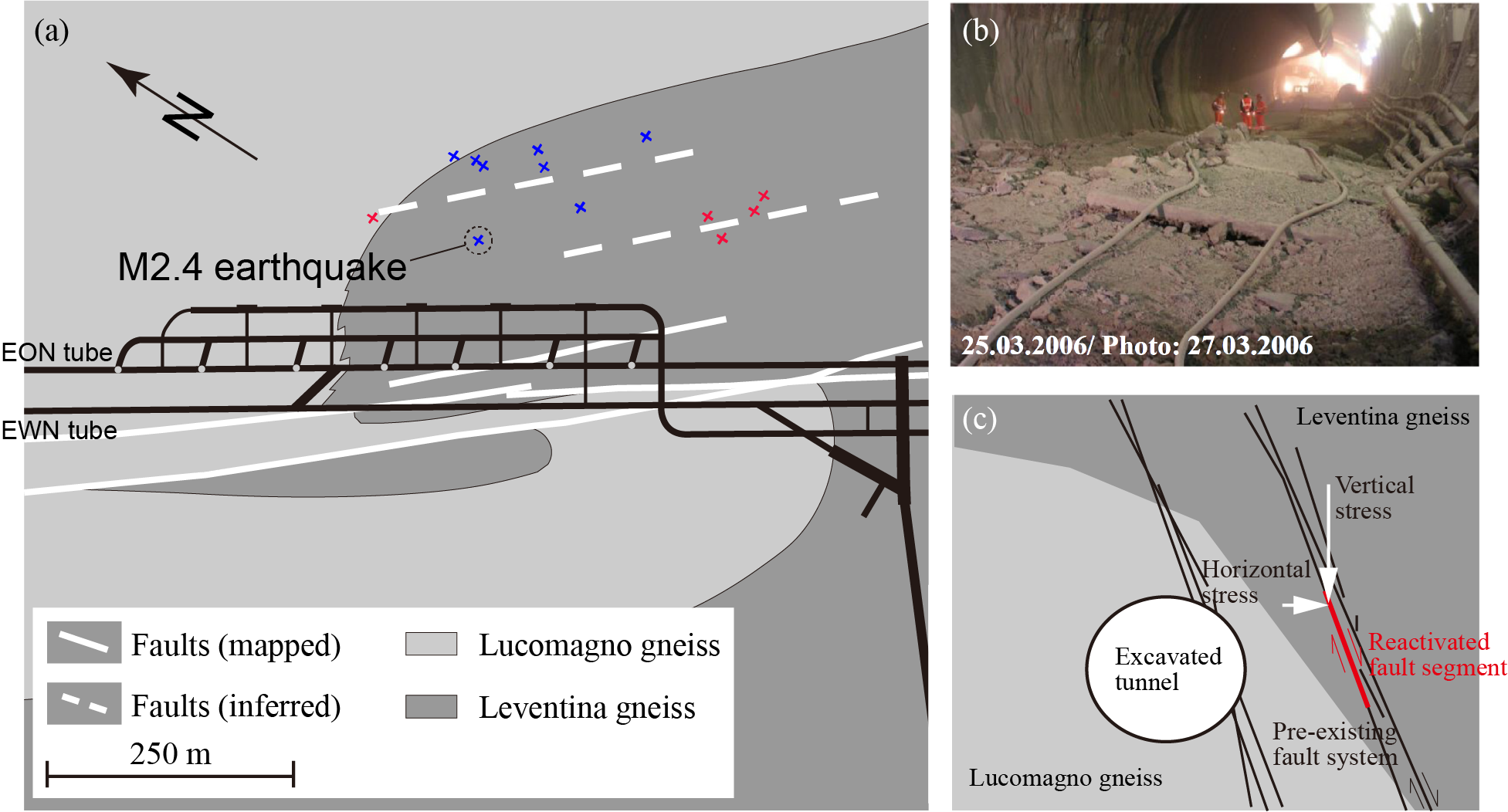}
	\caption{(a) Epicenter locations of earthquakes in two clusters, marked by blue and red crosses, located north of the Faido Multifunction Station \cite{husen2012induced}. The largest event (magnitude 2.4) triggered a rockburst within the Faido station. (b) Invert heave and bulking damage observed in the EON tunnel as a result of the rockburst \cite{rehbock2018fault}. (c) Conceptual fault model illustrating failure along a reactivated segment of the pre-existing fault system (highlighted in red), driven by increased differential stress associated with tunnel excavation \cite{husen2012induced}. In (a), solid white lines indicate mapped faults, while dashed white lines represent inferred faults based on fault-plane solutions and earthquake epicenter distributions. The invert damage directly caused by the M2.4 seismic event was relatively minor compared to what is displayed in (b), which was taken two days later and reflects the cumulative effects of this event along with several subsequent smaller events \cite{rehbock2018fault}. Note that both (a) and (c) are taken from the prior study\cite{husen2012induced} (Fig. 17 and Fig. 18b), respectively, and (b) is taken from the prior study\cite{rehbock2018fault} (Fig. 9).}
	\label{FIG:Gotthard_scenario}
\end{figure*}

The Gotthard Base Tunnel, spanning a total length of 57 km, serves as the centerpiece of the New Alpine Traverse through the Swiss Alps \cite{husen2012induced,sala2016gotthard,rehbock2018fault}. It comprises two parallel single-track tubes, denoted as EON and EWN (Fig. \ref{FIG:Gotthard_scenario}a), each with a diameter of 9.2 m. Two multifunction stations are located at approximately one-third and two-thirds of the tunnel length. During the construction of the Faido multifunction station, frequent and often severe rockbursts were reported from March 2004 \cite{rehbock2018fault}. To monitor seismic activity in this area, a dense network of seismic stations was installed between October and December 2005, and a total of 112 earthquakes were detected between October 2005 and August 2007. The largest of these events, with a magnitude of 2.4 (Fig. \ref{FIG:Gotthard_scenario}a), occurred on March 25, 2006. It caused substantial damage to the Faido station, including cracking in and significant spalling of the reinforced tunnel walls, as well as an uplift of around 0.5 m of the tunnel invert (Fig. \ref{FIG:Gotthard_scenario}b) \cite{husen2012induced,rehbock2018fault}. 

\begin{figure*}[htbp]
	\centering
	\includegraphics[width=1.0\textwidth]{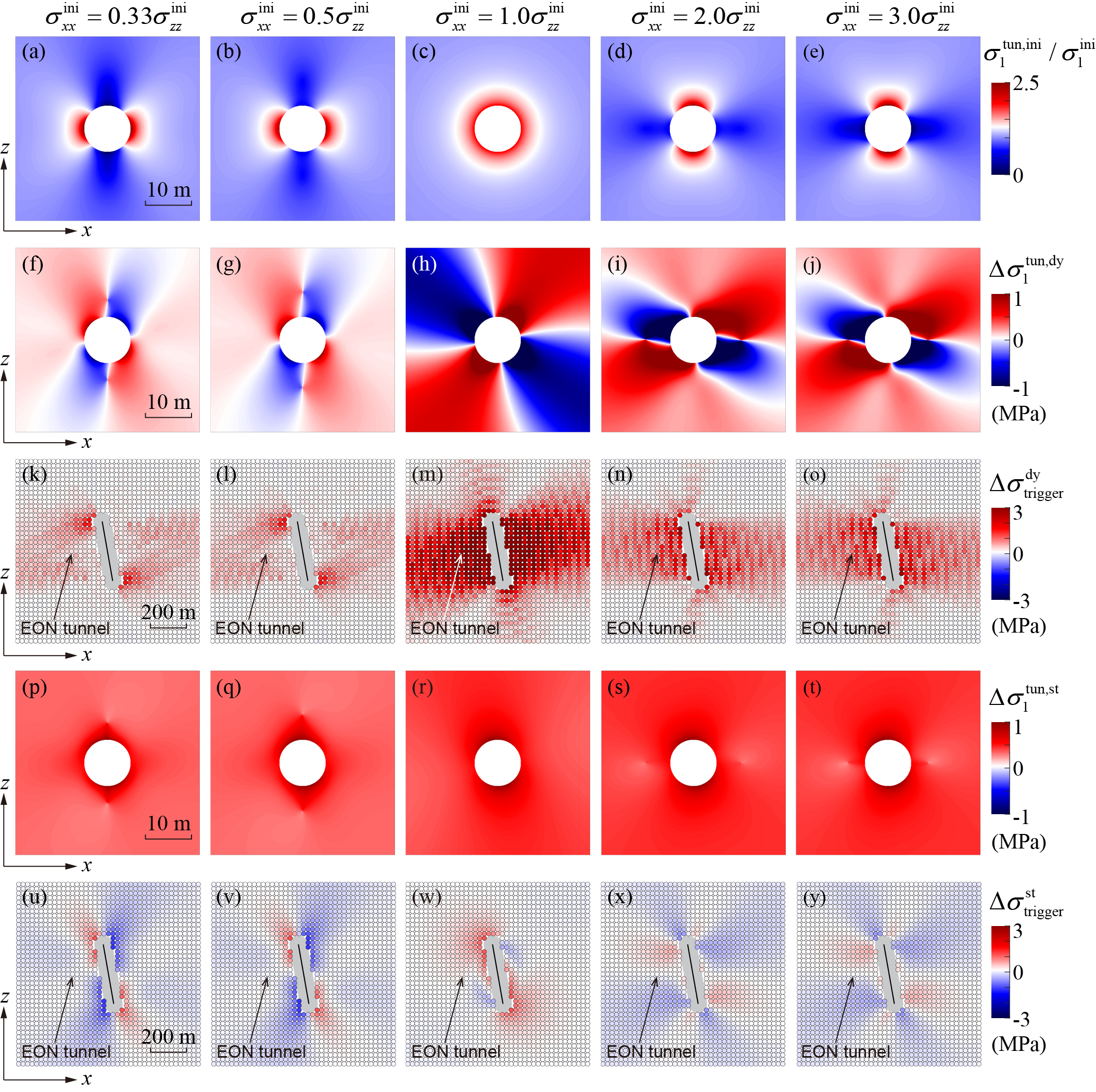}
	\caption{Static and dynamic triggering effects of the M2.4 seismic event on the EON tunnel at the Faido Multifunction Station. (a-e) Initial stress state around the tunnel, represented by $\sigma_1^{\rm{tun,ini}}$ normalized by $\sigma_1^{\rm{ini}}$, under different in-situ stress conditions. (f-j) and (p-t) Coseismic dynamic and static stress changes around the tunnel, indicated by $\Delta \sigma_1^{\rm{tun,dy}}$ and $\Delta \sigma_1^{\rm{tun,st}}$, respectively. (k-o) and (u-y) Dynamic and static triggering potentials for rockbursts, expressed by $\Delta \sigma_{\rm{trigger}}^{\rm{dy}}$ and $\Delta \sigma_{\rm{trigger}}^{\rm{st}}$, respectively. Results in panels (a-j) and (p-t) correspond to the tunnel located at the footwall, 200 m from the hypocenter, and at the same depth as the hypocenter. Panels (k-o) and (u-y) show results for a spatial matrix of tunnel locations, where the location of EON tunnel are denoted by arrows.}
	\label{FIG:Gotthard_results}
\end{figure*}

Husen et al. (2012) \cite{husen2012induced} identified the hypocenter location (Fig. \ref{FIG:Gotthard_scenario}a) and source dimensions of the M2.4 seismic event (Table \ref{tbl-faultslip-parameters}). Subsequent studies have explored the potential mechanisms underlying the reactivation of the fault segment responsible for this event \cite{husen2012induced,wu2017unloading,rinaldi2020fault,zhao2023causal}. However, the reason why such a small-magnitude earthquake produced severe rockburst damage in the EON tunnel, located about 200 m away from the hypocenter, remains unclear. The spatial relationship between the hypocenter and the EON tube is shown in Fig. \ref{FIG:Gotthard_scenario}a. To address this question, we apply our analytical model to the conceptual setup illustrated in Fig. \ref{FIG:Gotthard_scenario}c. Note that this schematic plot is not to scale and does not reflect the actual distance between the excavated tunnel and the reactivated fault segment. We evaluate the tunnel's responses under different in-situ stress conditions by varying the horizontal-to-vertical stress ratio, with the vertical stress assumed to be governed by the overburden effect. The analysis specifically considers a case where the tunnel is located 200 m away from the hypocenter and on the footwall side of the fault, at the same depth as the hypocenter. The model parameters are given in Table \ref{tbl-faultslip-parameters}, and the corresponding results are presented in Fig. \ref{FIG:Gotthard_results}.

\begin{table}[h]
\centering
\caption{Model parameters for the fault-slip rockburst analysis at Faido station.}
\label{tbl-faultslip-parameters}
\begin{tabularx}{0.95\linewidth}{@{} >{\raggedright\arraybackslash}X 
                                      >{\raggedright\arraybackslash}m{4.0cm} 
                                      >{\raggedright\arraybackslash}m{2.0cm} @{}}
\toprule
\textbf{Parameter} & \textbf{Value} & \textbf{Unit} \\
\midrule
\multicolumn{3}{@{}l}{\textit{Seismic source:}} \\
\quad Fault length $L$ & 336 & m \\
\quad Fault dip angle $\alpha$ & 80 & $^\circ$ \\
\quad Hypocenter depth $h_0$ & 1,240 & m \\
\quad Moment magnitude $M_{\mathrm{w}}$ & 2.4 & -- \\
\quad Seismic moment $M_{0}$ & $5.07 \times 10^{12}$ & N·m \\
\midrule
\multicolumn{3}{@{}l}{\textit{Rock mass:}} \\
\quad Young’s modulus $E$ & 35 & GPa \\
\quad Poisson’s ratio $\nu$ & 0.18 & -- \\
\quad Friction angle $\varphi$ & 36 & $^\circ$ \\
\quad Cohesion $c$ & 12.5 & MPa \\
\quad Uniaxial compressive strength $\sigma_{\rm{c}}$ & 49.07 & MPa \\
\bottomrule
\end{tabularx}

\vspace{0.5em}
\begin{minipage}{0.95\linewidth}
\footnotesize\textit{Note:} Seismic source parameters are adopted from \cite{husen2012induced}. Rock mass properties correspond to the Leventina gneiss and are reported in \cite{rehbock2018fault}. The uniaxial compressive strength is estimated as $\sigma_{\rm{c}} = 2c/\tan(45^\circ - \varphi/2)$ following \cite{lei2016characterisation}.
\end{minipage}
\end{table}

Prior to the M2.4 seismic event, the pattern of stress concentration around the tunnel is strongly controlled by the in-situ horizontal-to-vertical stress ratio, $\sigma_{xx}^{\rm{ini}}/\sigma_{zz}^{\rm{ini}}$ (Figs. \ref{FIG:Gotthard_results}a-e). When this ratio is less than one, stress concentrates at the tunnel sidewalls, while stress is released at the crown and invert. Conversely, when the ratio exceeds one, the crown and invert exhibit pronounced stress concentrations, whereas the sidewalls experience stress release. For a ratio equal to unity, stress is more uniformly distributed along the tunnel boundary. Following the M2.4 seismic event, both coseismic static and dynamic effects induce additional stress perturbations around the tunnel (Figs. \ref{FIG:Gotthard_results}f-j, p-t), leading to different potentials of rockburst triggering (Figs. \ref{FIG:Gotthard_results}k-o, u-y). Dynamic triggering effects are most pronounced for the case of  $\sigma_{xx}^{\rm{ini}}/\sigma_{zz}^{\rm{ini}}=1.0$, where the maximum tangential stress on the EON tunnel boundary increases by $\Delta \sigma_{\rm{trigger}}^{\rm{dy}}\approx3\ \rm{MPa}$. For stress ratios of 2.0 and 3.0, the corresponding stress increment is around 1 MPa. In contrast, for lower stress ratios of 0.33 and 0.5, the maximum tangential stress increases by only around 0.3 MPa. Static triggering effects, in comparison, induce relatively weak responses at the EON tunnel. For $\sigma_{xx}^{\rm{ini}}/\sigma_{zz}^{\rm{ini}}=1.0$, the maximum tangential stress increases by only $\Delta \sigma_{\rm{trigger}}^{\rm{st}}\approx0.2\ \rm{MPa}$, whereas for all other cases, the stress increment remains below 0.1 MPa.

These findings suggest that the rockburst event at the Faido station was likely triggered by dynamic effects associated with the M2.4 fault-slip event. The EON tunnel, where significant invert uplift and bulking were observed, is situated within the dynamic triggering-dominated region (Figs. \ref{FIG:Gotthard_results}m-o). Furthermore, in the $\sigma_{xx}^{\rm{ini}}/\sigma_{zz}^{\rm{ini}}=1.0$ case, the pre-earthquake maximum tangential stress around the tunnel reaches $\sigma_{\theta\theta}^{\rm{max,ini}} = 66.96\ \rm{MPa}$ (Fig. \ref{FIG:Gotthard_results}c), which exceeds the theoretical uniaxial compressive strength $\sigma_{\rm{c}}$ derived from site-specific rock strength parameters (see Table \ref{tbl-faultslip-parameters}). Other cases generate even higher values of $\sigma_{\theta\theta}^{\rm{max,ini}}$, indicating that the tunnel was already in a critically stressed state before the seismic event such that strainbursts may potentially occur during excavation. The additional coseismic dynamic stress increment, up to 3 MPa, may therefore have been sufficient to push the critically-stressed surrounding rock into failure, thereby initiating a fault-slip rockburst event \cite{zhu2010numerical,kaiser2013critical,he2017rock,su2017experimental,hu2018experimental}. Notably, in all three cases where dynamic triggering effects are substantial, the induced stress concentrations are most pronounced near the crown and invert of the tunnel (Figs. \ref{FIG:Gotthard_results}h-j), consistent with the field observations of severe invert damage in the EON tunnel (Fig. \ref{FIG:Gotthard_scenario}c).

This case study further demonstrates that the spatial distribution of stress concentration and release caused by tunnel excavation (Figs. \ref{FIG:Gotthard_results}a-e) can differ significantly from those induced by fault-slip events (Figs. \ref{FIG:Gotthard_results}f-j). Excavation-induced stress patterns are primarily controlled by the in-situ stress state, whereas the fault slip-induced responses are influenced by both the in-situ stress field and the tunnel's relative position to the fault, including its distance and orientation. As a result, the static and dynamic triggering effects (Figs. \ref{FIG:Gotthard_results}k-o, u-y) can vary substantially depending on the combined influence of the in-situ stress condition and the tunnel-fault configuration.
\FloatBarrier
\section{Discussion and conclusions}
\label{section:conclusions}
The occurrence of fault-slip rockbursts is primarily governed by two factors \cite{kaiser2013critical,manouchehrian2017analysis,su2017experimental,li2017failure,hu2018experimental,cai2020monitoring,hu2020rock,farhadian2021new,askaripour2022rockburst,he2023review,liu2024experimental,su2024influence}: (i) the pre-earthquake stress state in the rock mass surrounding the tunnel, and (ii) the static and dynamic stress changes induced by the seismic event at the tunnel. The pre-earthquake stress state arises from the transition of the surrounding rock mass from an initial triaxial stress state to a uniaxial or biaxial stress state due to tunnel excavation \cite{hu2018experimental,zhao2014influence,li2017failure,he2023review,gu2024effect}, resulting in the development of maximum tangential stress along the tunnel boundary\cite{cai2014situ,he2023review}, expressed as $\sigma_{\theta\theta}^{\rm{max,ini}}=3\sigma_1^{\rm{ini}}-\sigma_3^{\rm{ini}}$. The stability of the local rock mass, whether it remains stable, reaches a critically stressed state, or fails, depends on the relationship between $\sigma_{\theta\theta}^{\rm{max,ini}}$ and the uniaxial compressive strength $\sigma_{\rm{c}}$ \cite{dowding1986potential,hu2018experimental,he2023review,shi2025laboratory}. Subsequent seismic events associated with fault slip impose additional static and dynamic stress perturbations on the tunnel surroundings, which can either increase or decrease the local maximum tangential stress. These stress changes, denoted as $\Delta \sigma_{\rm{trigger}}^{\rm{st}}$ and $\Delta \sigma_{\rm{trigger}}^{\rm{dy}}$, can be analytically calculated using our model. The magnitudes of these stress increments depend not only on the tunnel's spatial relationship to the fault (see Fig. \ref{FIG:Triggering-role}), but also on the characteristics of the seismic source (Figs. \ref{FIG:Parameter_Sensivity}a-d), the mechanical properties of the surrounding rock mass (Figs. \ref{FIG:Parameter_Sensivity}e-h), and the in-situ stress state (Fig. \ref{FIG:Gotthard_results}). These rockburst triggering stress increments are essential for assessing whether a seismic event can drive the surrounding rock mass from a critically stressed state into catastrophic failure \cite{zhu2010numerical,kaiser2013critical,he2017rock,su2017experimental,hu2018experimental}, and also serve as the basis for classifying the triggering mechanism: static, dynamic, or dual triggering (Fig. \ref{FIG:Dominant-mechanisms}). Prior studies have employed various numerical and laboratory experiments to investigate whether rock failure can be triggered by superimposed dynamic stress pulses \cite{zhu2010numerical,he2012studies,wang2017numerical,hu2018experimental,su2017experimental,gao2021numerical,su2024influence}, thereby informing rockburst mechanisms and prediction. Our analytical framework can further offer a physically grounded and quantifiable approach to this problem. It provides both the time-dependent dynamic stress increment, expressed as $\sigma_{\theta\theta}^{\rm{max,dy}}(t)-\sigma_{\theta\theta}^{\rm{max,ini}}$ with a peak value of $\Delta \sigma_{\rm{trigger}}^{\rm{dy}}$, arising from seismic wave propagation, as well as the static stress increments $\Delta \sigma_{\rm{trigger}}^{\rm{st}}$ resulting from coseismic stress redistribution. Together, these components enable a more comprehensive assessment of fault-slip rockburst hazards. Furthermore, the analytical model facilitates the construction of spatial maps highlighting zones of elevated static and dynamic triggering potential, offering practical guidance for tunnel design and risk mitigation. This predictive capability supports proactive decision-making in identifying rockburst hazardous zones, implementing reinforcement strategies, and designing tunnels that are resilient to seismic hazards during deep underground construction or mining.

While the proposed analytical model provides valuable insights into the mechanisms of static and dynamic triggering of fault-slip rockbursts and serves a practical tool for rockburst hazard prediction, several limitations should be noted. First, the current study is based on a 2D framework, which limits its ability to capture the full complexity of 3D spatial relationships between the tunnel and the fault. As a result, scenarios involving tunnels positioned obliquely or noncoplanar to the seismogenic fault cannot be adequately represented. Second, time-dependent effects like subcritical crack growth due to stress corrosion are not considered in the current study. Such effects may contribute to rockbursts occurring under less critically stressed conditions and with delayed timing. Further work will aim to address these limitations by incorporating 3D modeling capabilities and time-dependent effects, thereby enhancing the predictive capability and applicability of the model to real-world geological/geotechnical conditions.

To conclude, we have developed a novel physics-based analytical framework to investigate the static and dynamic triggering mechanisms of fault-slip rockbursts in tunnels subjected to nearby or remote seismic fault rupture activity. The model captures coseismic static stress redistribution arising from shear stress drop associated with fault dislocation, as well as dynamic stress perturbations induced by seismic wave propagation. Together, these components enable a comprehensive assessment of rockburst risks in both near-field and far-field regions of a seismogenic fault. Our results reveal that both coseismic static and dynamic effects can trigger rockbursts. The nature and extent of triggering are governed by multiple factors, including the tunnel's spatial relationship to the fault (in terms of distance and orientation), the seismic source parameters, the mechanical properties of the surrounding rock mass, and the in-situ stress state. In far-field regions, dynamic triggering is the dominant mechanism, especially in directions perpendicular to the fault. In contrast, in near-field regions, a competition between static and dynamic triggering effects exists, with static triggering being more prominent in the fault's down-dip direction. To quantify triggering potential, we introduce two key stress-based indicators, $\Delta \sigma_{\rm{trigger}}^{\rm{st}}$ and $\Delta \sigma_{\rm{trigger}}^{\rm{dy}}$, which provide a physically grounded basis for evaluating whether coseismic stress changes are sufficient to exceed the local failure threshold, and for identifying rockburst mechanisms: static, dynamic, or dual triggering. The model further enables the generation of spatial maps that delineate zones of elevated static and dynamic triggering potential, thereby offering practical guidance for tunnel alignment, design optimization, and hazard mitigation in seismically active environments. To demonstrate the applicability of the model, we conducted a case study of a historical fault-slip rockburst induced by a M2.4 seismic event in the Gotthard Base Tunnel. The analysis indicates that dynamic triggering tends to be the dominant mechanism responsible for the observed rockburst event. Our findings advance the current understanding of fault-slip rockburst triggering mechanisms, and provide a parsimonious framework for assessing rockburst hazards in deep underground engineering applications.

\section*{Acknowledgments}
\label{sec:acknowledgements}
Z.Z. acknowledges this work was supported by the National Natural Science Foundation of China (Grant No. 42377146). Q.L. is grateful for the financial support by the Open Fund Project of the State Key Laboratory of Intelligent Coal Mining and Strata Control (Grant No. SKLIS202418).
\FloatBarrier
\appendix

\section*{Appendices}
\label{section:appendix}
\section{Examination of the analytical solution for coseismic static stress variation}
\label{appendix:static}
\begin{figure}[htbp]
	\centering
	\includegraphics[width=0.45\textwidth]{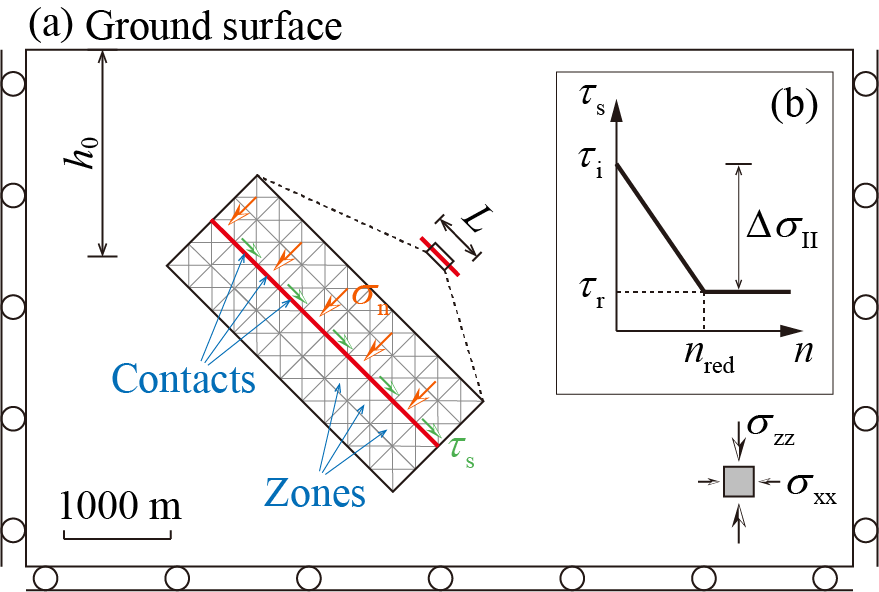}
	\caption{(a) Geometry of the numerical model, and (b) fault shear stress drop prescribed by a linear reduction function.}
	\label{FIG:numerical_model_static_validation}
\end{figure}

The analytical solution for coseismic static stress changes is examined via a comparison against numerical simulation results obtained using UDEC (Universal Distinct Element Code) \cite{itascauedc2019}. The benchmark is based on the fault-slip scenario described in Section \ref{Section:Fault-slip-description}. The geometry of the numerical model is illustrated in Fig. \ref{FIG:numerical_model_static_validation}, where the problem domain is discretized using a triangular finite-difference mesh. Element sizes range from 20 m near the fault to 80 m near the domain boundaries to ensure both accuracy and computational efficiency. The fault is represented as a series of contact interface elements, which enables the simulation of interaction forces between adjacent walls. The model is subjected to the initial stress condition defined by Eq. \ref{eq:initial_stress_condition}. Roller boundary conditions are applied along the lateral and bottom boundaries, whereas the top boundary is left free to represent the ground surface. 

We simulate the coseismic static stress redistribution induced by a shear stress drop of $\Delta \sigma_{\rm{II}}=5.84\ \rm{MPa}$. The shear stress acting on the fault plane, denoted as ${\tau _{\rm{s}}}$, is linearly reduced from its initial value $\tau_{\rm{i}}$ to a residual value $\tau_{\rm{r}}$ (where $\tau_{\rm{r}}=\tau_{\rm{i}}-\Delta \sigma_{\rm{II}}$ ) over 10,000 computational steps (see Fig. \ref{FIG:numerical_model_static_validation}b). During this process, the fault remains in a state of equilibrium between shear stress and shear strength, described by:
\begin{align}
    {\tau _{\rm{s}}} = \mu {\sigma _{\rm{n}}},
\end{align}
\noindent where $\sigma_{\rm{n}}$ is the normal stress acting on the fault plane, and $\mu$ is the friction coefficient. Since the normal stress $\sigma_{\rm{n}}$ remains nearly constant during the mode $\rm{II}$ failure process, the shear stress drop is simulated by gradually weakening the shear strength. This weakening is implemented by linearly reducing the friction coefficient according to:
\begin{align}
\mu = 
\begin{cases}
\mu_1-(\mu_1-\mu_2)\dfrac{n}{n_{\rm{red}}}, & n \leq n_{\rm{red}} \\[6pt]
\mu_2, & n > n_{\rm{red}}
\end{cases},
\label{eq:frictional_weakening}
\end{align}
\noindent where $n$ is the current computational step, $n_{\rm{red}}$ is the total number of reduction steps (10,000), $\mu_1 = \tau_{\rm{i}}/\sigma_{\rm{n}}$, and $\mu_2 = \tau_{\rm{r}}/\sigma_{\rm{n}}$. This slow and controlled reduction of shear strength is designed to suppress the generation of dynamic waves, thereby isolating the static stress response.

\begin{figure*}[htbp]
	\centering
	\includegraphics[width=1.0\textwidth]{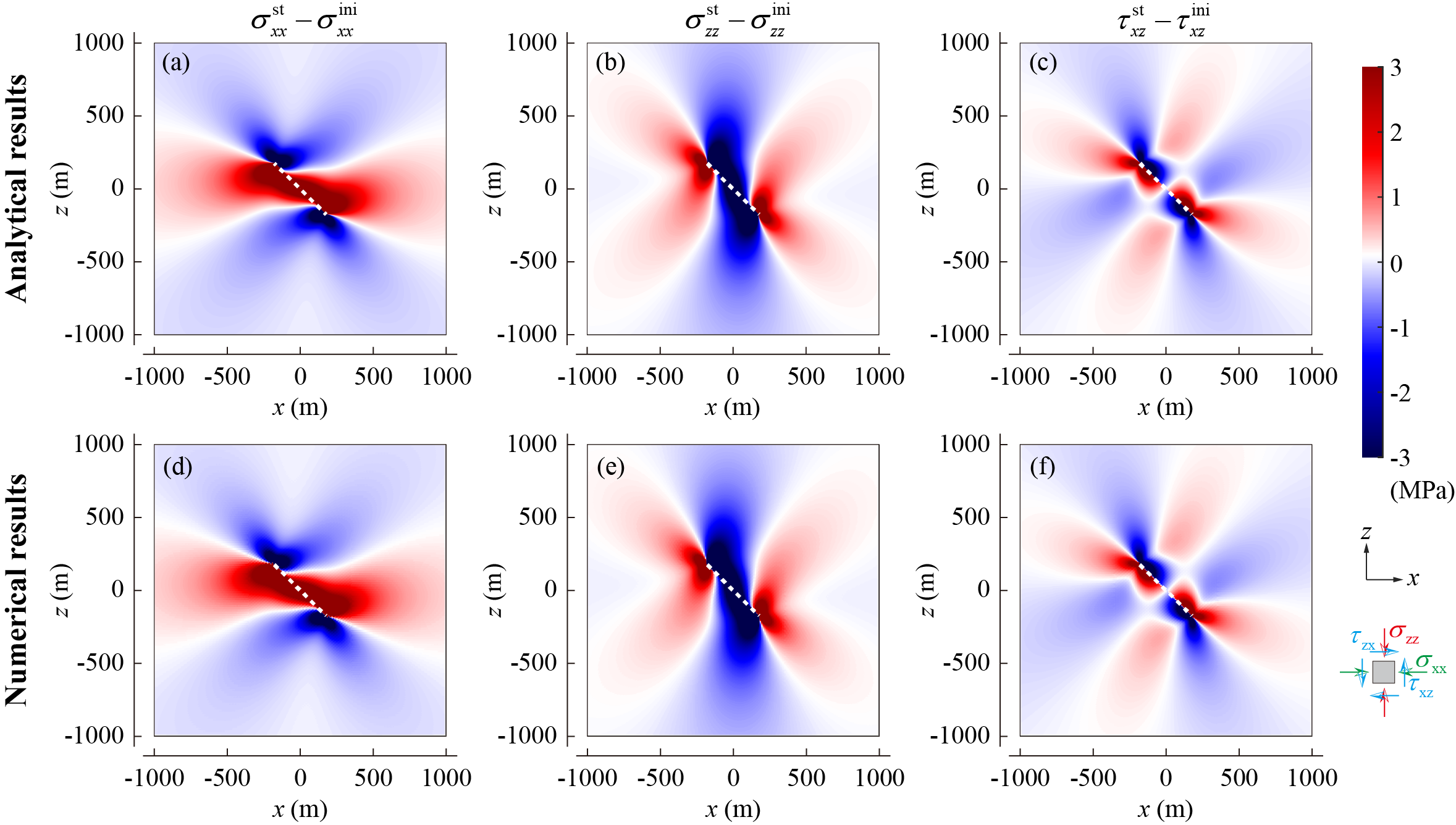}
	\caption{Coseismic static stress changes relative to the initial stress field, derived from (a-c) analytical solutions and (d-f) numerical simulations. The white dashed line indicates the location, orientation, and size of the seismogenic fault.}
	\label{FIG:static_validation}
\end{figure*}

\begin{figure*}[htbp]
	\centering
	\includegraphics[width=0.9\textwidth]{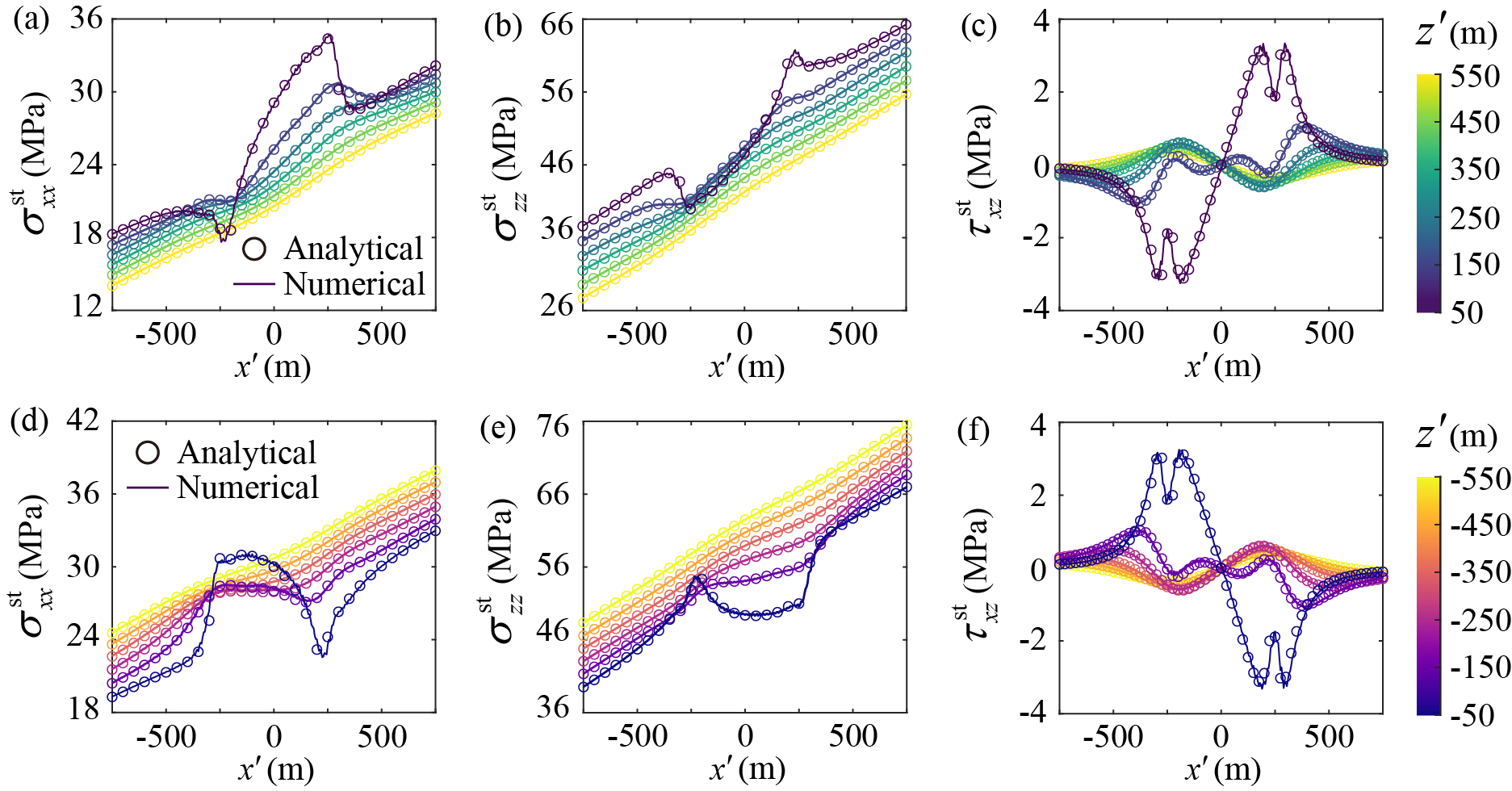}
	\caption{Comparison of coseismic static stress components between analytical solutions and numerical simulations: (a,d) $\sigma_{xx}^{\rm{st}}$, (b,e) $\sigma_{zz}^{\rm{st}}$, and (c,f) $\tau_{xz}^{\rm{st}}$. Stress values are extracted from a series of monitoring points distributed along the $x^{\prime}-z^{\prime}$ coordinate system (refer to Fig. \ref{FIG:SchematicModel}). Dots represent analytical solutions and solid lines denote numerical results. Both are color-coded by the corresponding $z^{\prime}$ coordinates of the monitoring points.}
	\label{FIG:stress_value_validation}
\end{figure*}

Fig. \ref{FIG:static_validation} presents the resulting changes in stress components relative to the initial state, including $\sigma_{xx}^{\rm{st}}-\sigma_{xx}^{\rm{ini}}$, $\sigma_{zz}^{\rm{st}}-\sigma_{zz}^{\rm{ini}}$, and $\tau_{xz}^{\rm{st}}-\tau_{xz}^{\rm{ini}}$. The numerical and analytical results show good agreement in both spatial distribution and magnitude. To quantitatively assess the accuracy, values of stress components ($\sigma_{xx}^{\rm{st}}$, $\sigma_{zz}^{\rm{st}}$, $\tau_{xz}^{\rm{st}}$) are further extracted from a series of monitoring points in both the numerical and analytical models. As shown in Fig. \ref{FIG:stress_value_validation}, the close match between the two sets of results confirms the validity of our analytical model in capturing coseismic static stress changes induced by fault slip.

\section{Examination of the analytical solution for coseismic dynamic stress perturbation}
\label{appendix:dynamic}
\begin{figure*}[htbp]
	\centering
	\includegraphics[width=1.0\textwidth]{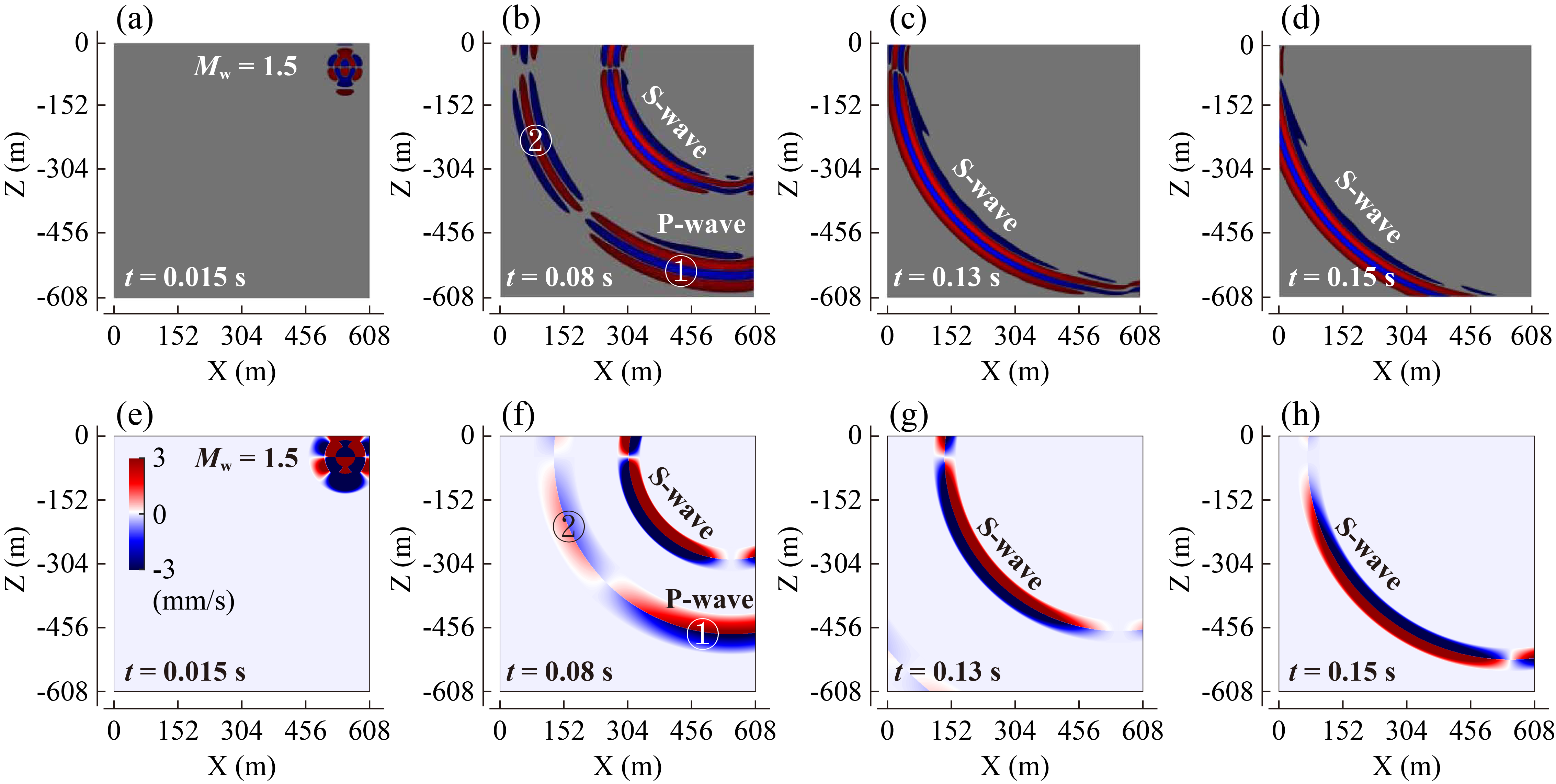}
	\caption{Snapshots of vertical velocity fields associated with dynamic wave propagation at 0.015, 0.08, 0.13, and 0.15 s for (a-d) the numerical results from \cite{wang2017numerical}, and (e-h) the corresponding analytical results obtained in this study. All results are presented in the coordinate system defined by Wang and Cai (2017) \cite{wang2017numerical}. Positive and negative vertical velocities, corresponding to upward and downward movements, are colored in red and blue, respectively. Symbols \ding{172} and \ding{173} indicate different populations of P-waves with different waveform characteristics.}
	\label{FIG:dynamic_validation_contour}
\end{figure*}

The analytical solution for coseismic dynamic stress perturbation is examined via a comparison against a numerical simulation of seismic wave propagation generated by fault slip, conducted by Wang and Cai (2017) \cite{wang2017numerical} using the SPECFEM2D software package. In their study, the seismic source corresponds to a normal fault-slip event with a moment magnitude of $M_{\rm{w}}=1.5$, a dip angle of $45^{\circ}$, and a hypocenter located at $\rm{X} = \rm{550\ m}$, $\rm{Z = -50\ m}$ (refer to the coordinate system in Fig. \ref{FIG:dynamic_validation_contour}). The rock mass is characterized by P-wave and S-wave velocities of 5,874 m/s and 3,400 m/s, respectively. Snapshots of the vertical velocity field from the numerical simulation are shown in Figs. \ref{FIG:dynamic_validation_contour}a-\ref{FIG:dynamic_validation_contour}d. Using the same source parameters, we obtain analytical results based on the formulation described in Section \ref{section:coseismic_dynamic_solution}, as shown in Figs. \ref{FIG:dynamic_validation_contour}e-\ref{FIG:dynamic_validation_contour}f.

The analytical and numerical results show good agreement in the overall propagation characteristics of the seismic waves. Both methods clearly distinguish between P-waves and S-waves. Notably, both identify that the P-wave field can be divided into different populations based on waveform characteristics, with a transition at approximately $45^{\circ}$ from the $\rm{Z}$-axis. The population labeled as \ding{172} displays a higher amplitude than that labeled as \ding{173} (see Figs. \ref{FIG:dynamic_validation_contour}b and \ref{FIG:dynamic_validation_contour}f). However, the numerical simulation produces more complex waveforms, with multiple polarity reversals (e.g., a "$+-+-$" pattern, where "$+$" and "$-$" denote the positive and negative vertical velocities, respectively), whereas the analytical solution yields simpler waveforms, typically showing a single polarity reversal (e.g., a "$-+$" pattern). This discrepancy primarily results from the different source time functions adopted. The numerical model uses a Ricker wavelet with a dominant frequency of 100 Hz, whereas our analytical model employs a source time function derived from seismic observations, which reflects the physics-based, self-similar growth process of fault rupture.

\begin{figure}[htbp]
	\centering
	\includegraphics[width=0.45\textwidth]{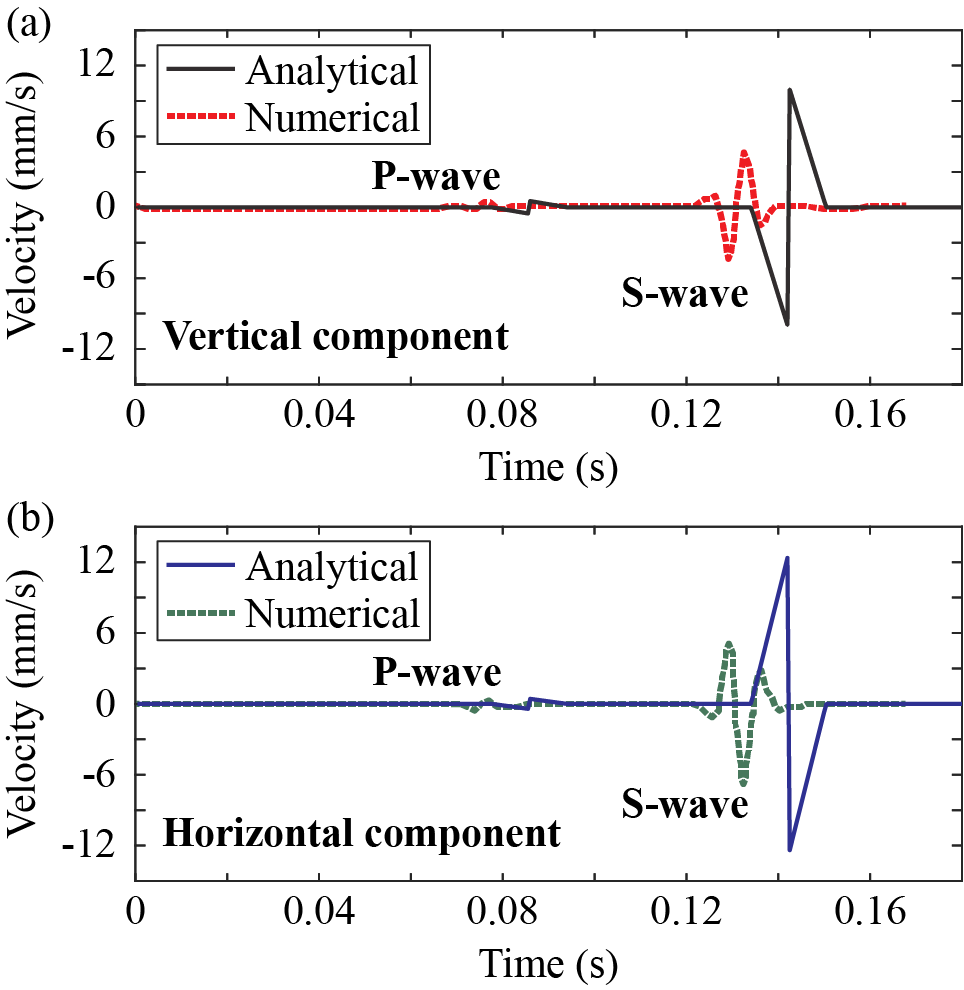}
	\caption{Coseismic dynamic wave velocities for (a) the vertical component and (b) the horizontal component. The results from our analytical model are shown as dashed lines, while the numerical results from Wang and Cai (2017) \cite{wang2017numerical} are shown as solid lines. The velocity histories are recorded at a receiver located at X = 264 m and Z = -405 m.}
	\label{FIG:dynamic_validation_curve}
\end{figure}

We further compare the time histories of velocity components recorded at a receiver positioned at $\rm{X=264\ m}$, $\rm{Z=-405\ m}$, as shown in Fig. \ref{FIG:dynamic_validation_curve}. Both the numerical and analytical results produce similar waveforms in both vertical and horizontal components. The observed amplitude differences are attributed to two factors: (1) the different source time function used, and (2) the distinct attenuation treatments in the two approaches. In our model, wave attenuation is incorporated analytically in Eq. \ref{eq:dynamic_displace}, whereas in the numerical simulation it is governed by user-defined damping parameters. It is also noted that in the analytical model, the arrival times of the P- and S- waves are slightly delayed compared to the numerical results, a discrepancy also visible in the wavefield snapshots (Fig. \ref{FIG:dynamic_validation_contour}). Nevertheless, the arrival times in the analytical results appear to be more consistent with the specified P- and S-wave velocities, i.e., $V_{\rm{P}}=5,874\ \rm{m/s}$, $V_{\rm{S}}=3,400\ \rm{m/s}$.

\section{Influence of fault discretization as an assembly of point sources}
\label{appendix:determination-N}
\begin{figure}[htbp]
	\centering
	\includegraphics[width=0.45\textwidth]{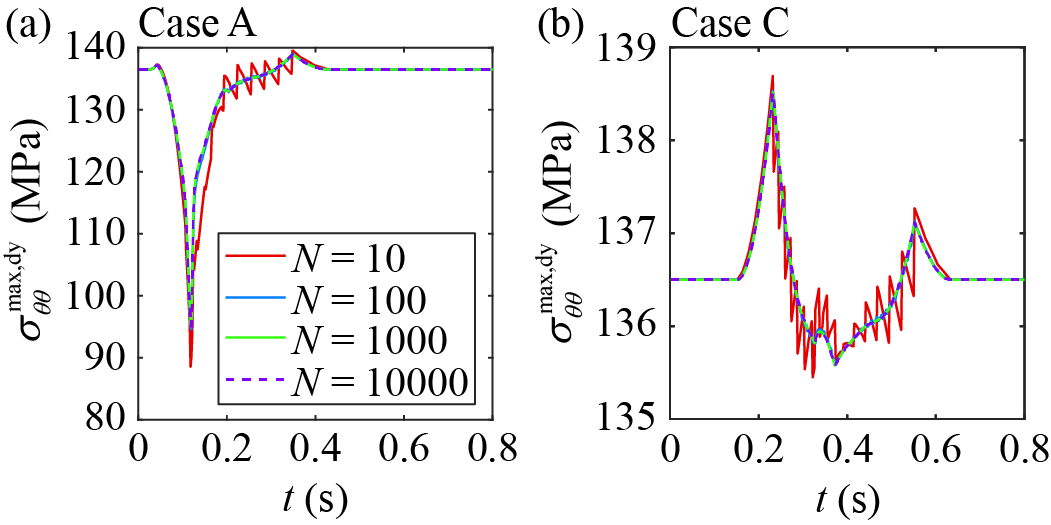}
	\caption{Temporal evolution of the coseismic dynamic maximum tangential stress $\sigma_{\theta\theta}^{\rm{max,dy}}$ along the tunnel boundary in (a) Case A and (b) Case C. Results are shown for fault discretizations using 10, 100, 1000, and 10000 point sources.}
	\label{FIG:test_discretize_number}
\end{figure}

To assess the influence of the discretization number $N$ (defining the number of fault segments or point sources) on the computed coseismic response, we calculate the dynamic maximum tangential stress, $\sigma_{\theta\theta}^{\rm{max,dy}}$ (see Table \ref{tbl-symbols} for definition), at tunnel locations A and C (refer to Fig. \ref{FIG:Dynamic_Stress_Changes_in_large_scale} for their positions). The results, shown in Fig. \ref{FIG:test_discretize_number}, indicate that the values of $\sigma_{\theta\theta}^{\rm{max,dy}}$ converge as $N$ exceeds 100. This suggests that the discretization number adopted in the paper ($N=\rm{1,000}$; see Section \ref{Section:Fault-slip-description}) is sufficiently high to ensure numerical accuracy and stability.

\FloatBarrier

\printcredits

%% Loading bibliography style file
% \bibliographystyle{model1-num-names}
\bibliographystyle{elsarticle-num}

% Loading bibliography database
\bibliography{cas-refs}

\end{document}